\documentclass{article}

\usepackage{PRIMEarxiv}

\usepackage[utf8]{inputenc} 
\usepackage[T1]{fontenc}    
\usepackage{hyperref}       
\usepackage{url}            
\usepackage{booktabs}       
\usepackage{amsfonts}       
\usepackage{nicefrac}       
\usepackage{microtype}      
\usepackage{lipsum}
\usepackage{fancyhdr}       
\usepackage{graphicx}       
\graphicspath{{media/}}     
\usepackage{tabularx}
\usepackage{graphicx}
\usepackage{arydshln}
\usepackage{microtype}
\usepackage{siunitx}
\usepackage{amsmath, nccmath}
\usepackage{cite}
\usepackage{enumitem}
\usepackage{mathtools}
\usepackage{multirow}
\usepackage{adjustbox}
\usepackage{bm}
\usepackage{wasysym}
\usepackage{amssymb}

\title{High Frequency Response of Volatile Memristors
}

\author{
  Ioannis Messaris, Alon Ascoli, Ahmet S. Demirkol, Vasileios Ntinas, Dimitrios Prousalis, and Ronald Tetzlaff \\
  Faculty of Electrical and Computer Engineering  \\
  Institute of Circuits and Systems \\
  Dresden, Germany\\
  \texttt{ioannis.messaris@tu-dresden.de} \\
}

\begin{document}
\maketitle

\begin{abstract}
In this theoretical study, we focus on the high-frequency response of the electrothermal $\textrm{NbO}_2$-Mott threshold switch, a real-world electronic device, which has been proved to be relevant in several applications and is classified as a volatile memristor. Memristors of this kind, have been shown to exhibit distinctive non-linear behaviors crucial for cutting-edge neuromorphic circuits. In accordance with well-established models for these devices, their resistances depend on their body temperatures, which evolve over time following Newton's Law of Cooling. Here, we demonstrate that HP's $\textrm{NbO}_2$-Mott memristor can manifest up to three distinct steady-state oscillatory behaviors under a suitable high-frequency periodic voltage input, showcasing increased versatility despite its volatile nature. Additionally, when subjected to a high-frequency periodic voltage signal, the device body temperature oscillates with a negligible peak-to-peak amplitude. Since, the temperature remains almost constant over an input cycle, the devices under study behave as linear resistors during each input cycle. Based on these insights, this paper presents analytical equations characterizing the response of the $\textrm{NbO}_2$-Mott memristor to high-frequency voltage inputs, demarcating regions in the state space where distinct initial conditions lead to various asymptotic oscillatory behaviors. Importantly, the mathematical methods introduced in this manuscript are applicable to any volatile electrothermal resistive switch. Additionally, this paper presents analytical equations that accurately reproduce the temperature time-waveform of the studied device during both its transient and steady-state phases when subjected to a zero-mean sinusoidal voltage input oscillating in the high-frequency limit. This analytical approach not only increases our comprehension of volatile electrothermal memristors but also provides a theoretical framework to harness the enhanced dynamical capabilities of real-world volatile memristors, potentially paving the way for advancements in smart system applications.
\end{abstract}


\section{Introduction}
\label{sec1}
As the development of AI, Cyber-Physical Systems, and Internet of Things devices continues apace, there is a growing need for groundbreaking progress in applications such as memory, sensing, in-memory computing, neuromorphic computing, and machine learning \cite{Song2023}. The non-volatile memristor has emerged as a promising 2-terminal electronic device to enable unlocking the full potential of such smart systems mainly due to its ability to store and process data within a single nanoscale volume. Contrary to a non-volatile memristor, which preserves its programmed resistance even when powered off, the resistance of a volatile memristor resets to a default value after power-off, regardless of any previously applied signals across its terminals. Mott threshold switches based on $\textrm{VO}_2$ \cite{Brown2022,Demirkol2022a}, ovonic threshold switch (OTS) devices \cite{Hennen2018}, and $\textrm{NbO}_2$ nanostructures \cite{9629238,Ascoli2021a,Slesazeck2015,Pickett2012} represent typical examples of volatile memristors, which are notable for featuring a distinctive negative differential resistance (NDR) region in their DC current-voltage plots. In fact, the unique non-linear phenomena, appearing across their nanoscale physical media, have established volatile memristors as critical components in state-of-the-art neuromorphic circuits, serving to mimic specific neuronal activities, including action potential \cite{Kumar2020}, as well as all-or-non spiking activity \cite{Pickett2013,Yi2018}, and plasticity \cite{Wang2018}, among other functions. Moreover, volatile memristors are locally-active devices, and the insights presented in \cite{9689062} and \cite{9855410} exemplify their application in understanding the intricacies of biological systems.

The highly non-linear behaviors of volatile and non-volatile memristors make them particularly attractive for circuit-theoretic investigations since they have been shown to exhibit intricate phenomena under bespoke stimuli, unmet in conventional electronic devices. Such is the fading memory phenomenon, according to which the steady-state response of a memristor to an input from a certain class depends on the input characteristics, rather than the initial conditions of the memristor itself. This memristor behavior was first identified in the study by Ascoli \textit{et al.} in \cite{Ascoli2016Erase} based on experimental data obtained from a non-volatile nanoscale $\textrm{TaOx}$-based RRAM device fabricated \cite{Yang2010} and modeled \cite{Strachan2013} at Hewlett Packard (HP) Labs. In this study, the fading memory effect emerged when the RRAM device was subjected to a suitable AC voltage input and was further analyzed from a circuit-theoretic perspective. A clarifying mathematical treatment of this phenomenon was provided in \cite{9956789}, verifying a hypothesis suggested in \cite{Ascoli2018, AscoliAdv}, according to which the asymmetry in the on- and off-switching kinetics in non-volatile memristors, which is ubiquitous in real-world devices of this kind, acts as the main mechanism underpinning its emergence. Having established a better understanding of the mechanisms underlying the emergence of fading memory in non-volatile memristors and identified the main factors that modulate this nonlinear phenomenon, a theoretical method for programming HP's $\textrm{TaOx}$ resistive switch to any target state was proposed in \cite{9956789}. This programming scheme involved the proper configuration of the DC offset level of a high-frequency square-wave AC periodic voltage input at the MHz frequency limit serving as a promising application example exploiting fading memory in memristors. The method introduced in \cite{9956789} was based on the study of the time evolution of the mean state value of the HP memristor by applying a time averaging-based approximation of its model, which was initially proposed by Pershin and Slipko \cite{Pershin2019}. In this work, the two authors introduced a graphic method, currently referred to as the Time Average State Dynamic Route (TA-SDR) \cite{Ascoli2023b}, which can be used to study first-order memristors under periodic square-wave-based stimuli. Specifically, the TA-SDR may enable the prediction of the asymptotic oscillatory behavior of the memory state, starting from a predefined initial condition. Further developing the research presented in \cite{9956789} and \cite{Pershin2019}, Ascoli \textit{et al.} \cite{Ascoli2023} introduced a novel analytical tool coined as the State Change Per Cycle Map (SCPCM), a first-order discrete-time map, which simplifies the exploration of first-order non-autonomous continuous-time systems by illustrating how the stimulus configures each admissible memory state of the device during a single input cycle. By utilizing the SCPCM on HP's $\textrm{TaOx}$ memristor model, the authors in \cite{Ascoli2023} revealed, for the first time, the potential of non-volatile memristors to exhibit multistability, as an expression of local fading memory, when stimulated by trains with a suitable number of positive and negative voltage pulses per period. In an input-induced multistable memristor, each locally stable steady-state response has a well-defined basin of attraction in the state space, leading to a unique asymptotic behavior regardless of the initial condition. Importantly, the study in \cite{Ascoli2023} highlighted that to fully harness the potential of memristors, circuit designers may consider going beyond merely exciting these devices with standard AC and DC inputs, exploring more complex stimuli to fully exploit the dynamical richness that arises from the inherent complexity of the physical processes underlying their operation. In this context, volatile memristors appear more complex than their non-volatile counterparts. As shown in \cite{7557034,Ascoli2016}, volatile memristors can exhibit a couple of dynamical regimes even under purely-AC sinusoidal inputs, as opposed to the more intricate stimulation schemes necessary to observe bistability in non-volatile memristors, as highlighted in \cite{Ascoli2023}. However, no study to date has analytically explained the mechanisms driving local fading memory in volatile real-world memristor devices. This will be the primary focus of the present study.  

The analyses presented in this paper will revolve around the $\textrm{NbO}_2$-Mott threshold switch, which is possibly the dynamically richest nanodevice falling into the class of volatile memristors. This device has been the subject of extensive research in numerous recent publications \cite{9629238,Kumar2020,Kumar2018,Kumar2017b,Kumar2017a,Gibson2016,Demirkol2022,Nandi2019,Li2019,Kumar2017}. This memristor may exhibit three locally stable steady-state dynamical behaviors under a suitable DC voltage bias stimulus \cite{Kumar2018}, which have been found to be contingent upon its body temperature, that is the dominant state variable in this device. Thus, with an appropriate constant voltage level applied across the device terminals, three basins of attraction may emerge, manifested as three contiguous body temperature ranges, where each basin of attraction leads to a distinct asymptotic steady-state device response, regardless of the initial body temperature. Volatile $\textrm{VO}_2$-based memristors and OTS devices can also display similar behaviors. However, these memristors typically display up to two steady-state responses when subjected to a DC voltage input. 
It's worth noting that several recently published physics-based compact models for these devices \cite{Brown2022, Demirkol2022a, 9629238, 9181036,Ascoli2021a} have successfully captured their circuit responses using relatively simple mathematical expressions. In this paper, we leverage the established predictive capability of the $\textrm{NbO}_2$-Mott memristor model, presented in \cite{9629238,9181036}, to analytically study the fading memory phenomenon in practical volatile threshold switches excited by zero-mean periodic voltage inputs. Even though the device under investigation is the volatile $\textrm{NbO}_2$-Mott memristor, it's important to emphasize that all the assumptions, mathematical analyses, and scientific findings presented here can be straightforwardly applied to any electro-thermally driven volatile memristor device.

Our analysis will primarily target the high-frequency limit, where the volatile memristor exhibits the richest dynamical behaviors under AC inputs, particularly within the context of fading memory. Specifically, we will demonstrate that under purely-AC voltage stimulation, as the frequency of the AC input increases, the $\textrm{NbO}_2$-Mott nanodevice initially exhibits a single globally stable steady-state behavior. As the input frequency further increases, it undergoes a bifurcation into two distinct behaviors, eventually reaching a point where it displays three locally stable steady-states, and thus three basins of attraction. Based on the models in \cite{Brown2022, Demirkol2022a, 9629238, Ascoli2021a} volatile nano-electronic threshold switches are equivalent to linear passive resistors under high-frequency excitation. Here, we present a theoretical approach that harnesses this characteristic behavior to formulate analytical equations capable of accurately predicting the transient and steady-state response of the $\textrm{NbO}_2$-Mott nanodevice under any sufficiently high-frequency periodic input, and for any initial condition in each of the three admissible basins of attraction. Our methodology is based on the behavioral observation that the net change in the device resistance over each cycle of a zero-mean AC periodic input is negligible, when the input frequency is sufficiently high \cite{9956789}. Under these conditions, the input power over each cycle of the high-frequency purely-AC periodic input waveform corresponds to its effective RMS value, further simplifying the analysis of the memristor response. Let us initiate our investigations by briefly presenting the DAE set model of the volatile memristor device under study. 

\section{The $\textrm{NbO}_2$-Mott Memristor Model}
\label{sec2}

\begin{figure}[t]
\centering
\includegraphics[width=0.5\linewidth]{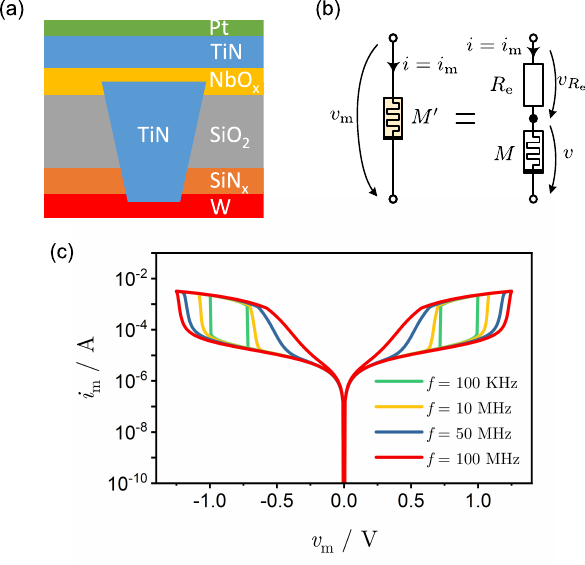}
\caption{Panel (a) visualizes a cross section of a sample $\textrm{NbO}_2$-Mott nano-device manufactured at HP Labs. Panel (b) draws the equivalent circuit schematic of the device shown in (a). Panel (c) illustrates simulated steady-state current-voltage responses of $M'$ under four sinusoidal voltage inputs that share the same amplitude, but differ in frequency (see body text for further details).}    
\label{Fig1}
\end{figure}

Panel (a), in Fig. \ref{Fig1}, illustrates a cross-section of a $\textrm{NbO}_2$-Mott memristor sample manufactured at HP Labs. Panel (b) draws its equivalent circuit diagram, where the series resistance $R_\textrm{e}$ is used to model the device $\textrm{Pt}$ and $\textrm{W}$ electrode resistances. The authors of the works presented in \cite{Kumar2017b,Gibson2016,Kumar2017a} showed that current-induced joule heating is the main physical mechanism affecting the device resistance and, thus, concluded that the most efficient approach to model the rate at which the device body temperature $T$ changes in terms of the input power $p$ dissipated in the device and of the heat transferred to the environment, was through Newton's Law of cooling, which is expressed as
\begin{equation}
\label{NbOx}
\frac{dT}{dt}=\frac{p}{C_\textrm{th}}-\frac{T-T_\textrm{amb}}{R_\textrm{th}(T)C_\textrm{th}},
\end{equation}
where
\begin{equation}
\label{power}
p=i\cdot v,
\end{equation}
and
\begin{equation}
\label{Rth}
R_{\textrm{th}}(T)=
\begin{cases} 
      R_{\textrm{th,ins}},\,\,\,T\leq T_C \\
      R_{\textrm{th,met}},\,\,\, T>T_C. \\
\end{cases}
\end{equation}
Variables $i$ and $v$ are the memristor core $M$ current and voltage, respectively, as shown in Fig. \ref{Fig1}(b). Function $R_\textrm{th}(T)$ is used to model the sharp increase in the device thermal resistance when it undergoes the Mott metal-insulator transition (MIT) at $T=T_C=\SI{1070}{K}$ \cite{Kumar2017b}. Thus, $R_\textrm{th,ins}=1.27\cdot 10^{-6}\,\textrm{K}\cdot \textrm{W}^{-1}$ and $R_\textrm{th,met}=1.91\cdot 10^{-6}\,\textrm{K}\cdot \textrm{W}^{-1}$ are the device thermal resistances in the insulating and metallic states, respectively. Parameter $C_\textrm{th}=10^{-16}\,\textrm{W}\cdot \textrm{s}\cdot \textrm{K}^{-1}$ is the thermal capacitance of the device, while $T_\textrm{amb}=\SI{300}{K}$ is the ambient temperature. On the other hand, the works in \cite{9629238} and \cite{9181036} revealed that the unique dynamical features of this nano-device can be accurately reproduced by coupling \eqref{NbOx}-\eqref{Rth} with a relatively simple Ohm's Law expression, in which the device memductance $G(\cdot)$ depends solely on the device body temperature $T$. Specifically,
\begin{figure}[t]
\centering
\includegraphics[width=0.55\linewidth]{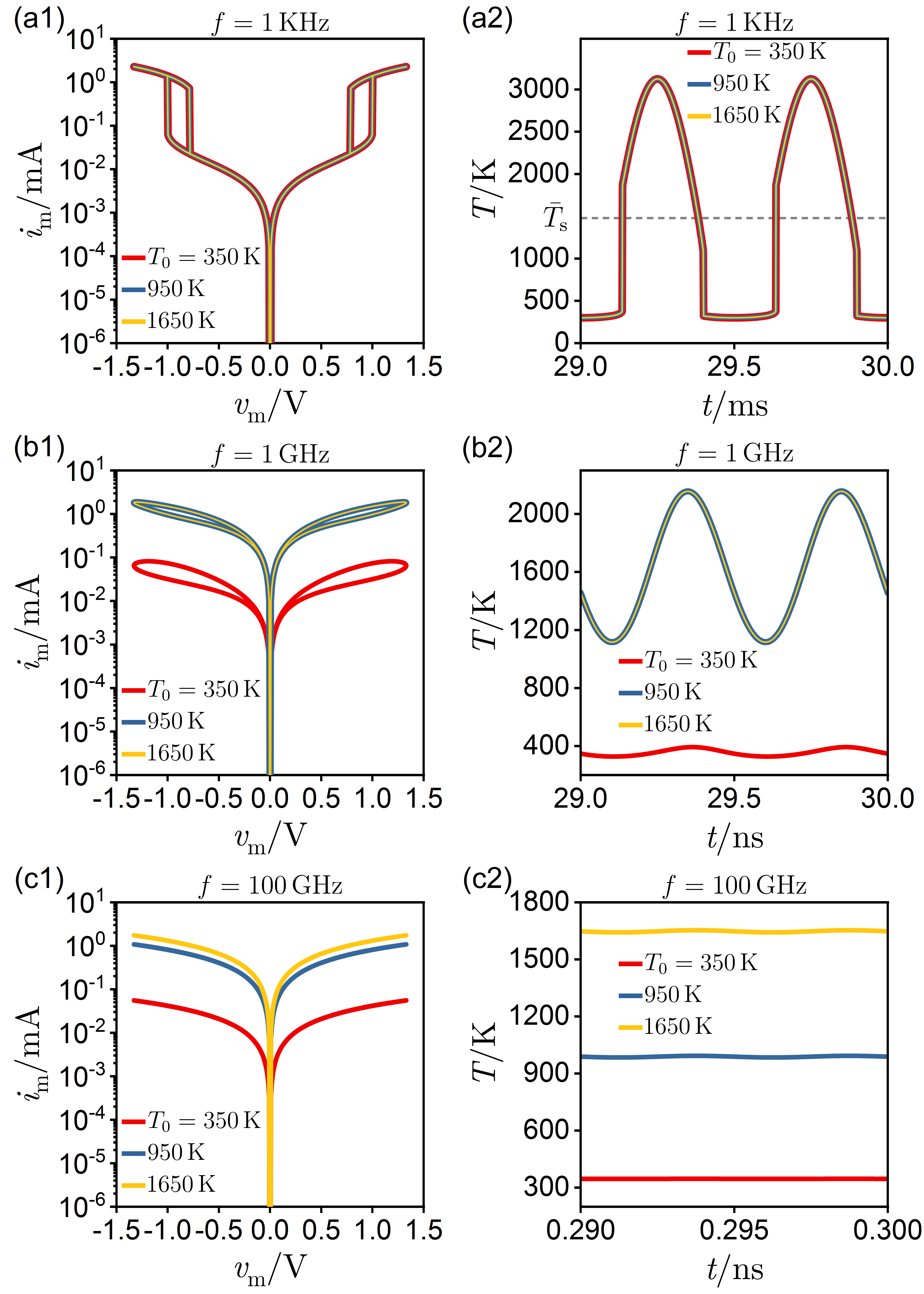}
\caption{Simulated steady-state current-voltage responses (see panels (a1), (b1), and (c1)) and steady-state temperature time-waveforms (see panels (a2), (b2), and (c2)) for the $\textrm{NbO}_2$-Mott memristor. The visualized plots were obtained by applying zero-mean sinusoidal voltage inputs across the terminals of $M'$ (see Fig. \ref{Fig1}(b)) at frequencies of $\SI{1}{KHz}$, $\SI{1}{GHz}$, and $\SI{100}{GHz}$, with an amplitude $\hat{v}_\textrm{m} = \SI{1.33}{V}$. Each panel displays three curves representing different initial conditions $T_0$ for the device body temperature $T$. Specifically, for the red, blue, and yellow curves, $T_0$ was set to $\SI{350}{K}$, $\SI{950}{K}$, and $\SI{1650}{K}$, respectively.}    
\label{Fig2}
\end{figure}
\begin{equation}
\label{im}
i=G(T)\cdot v
\end{equation}
where,
\begin{equation}
\label{conductancesimple}
G(T)=G_0\cdot e^{-\frac{g\cdot E_a}{2\cdot k_\textrm{B}\cdot T}},
\end{equation}
with 
\begin{equation}
\label{conductancesimple0}
G_0=\frac{\sigma_0\cdot(2\cdot a +1)\cdot c \cdot A}{2\cdot d}.
\end{equation}
Parameters $A=5.02\cdot10^{-15}\,\textrm{m}^2$, $d=8\cdot 10^{-9}\,\textrm{m}$, and $k_\textrm{B}=8.62\cdot 10^{-5}\,\textrm{eV}\cdot \textrm{K}^{-1}$, are the device lateral area, the $\textrm{NbO}_2$ film thickness, and the Boltzmann constant, respectively. Parameters $\sigma_0=1.55\cdot10^{-15}\,\textrm{S}\cdot \textrm{m}$ and $E_a=\SI{0.301}{eV}$ are material constants physically described in \cite{Gibson2016}. Finally, $a_0=0.123$, $g=0.986$, and $c=1.008$ are constants, that were used to improve the capability of the model to fit physically-obtained experimental data \cite{9629238,9181036}. 
We should note here, that, based on memristor theory, equations \eqref{NbOx}-\eqref{conductancesimple} classify the $\textrm{NbO}_2$-Mott memristor core $M$ as a first-order \textit{generic} memristor \cite{Chua2015}\footnote{According to \cite{Chua2015}, the high level mathematical representation of a first-order voltage-controlled \textit{generic} memristor is described by a differential algebraic equation (DAE) set model, where the Ohm's Law and state equation are expressed as, $i=G(x)\cdot v$ and $\dot{x}=f(x,v)$, respectively. $G(\cdot)$ is the memductance function, while $f(\cdot)$ captures the time-evolution of the device state $x$. Typically, the state variable in electro-thermal memristors, such as the $\textrm{NbO}_2$-Mott nano-device under study, is chosen as the device body temperature $T$. Notice, that by substituting \eqref{im} in \eqref{NbOx}, the time-derivative of the memristor core $M$ body temperature $\dot{T}$ can be expressed as being dependent on the temperature $T$ itself as well as on the voltage drop across the memristor core $v$ (see Fig. \ref{Fig1}(b)).}. As already shown in \cite{Demirkol2022}, when a linear resistance is combined in series with a \textit{generic} memristor, the resulting one-port is also a \textit{generic} memristor. This conclusion can be easily demonstrated by using basic algebraic techniques. Hence, the device formed by connecting $R_\textrm{e}$ and $M$ in series, denoted as $M'$ (refer to Fig. \ref{Fig1}(b)), can be classified as a \textit{generic} memristor as well. Its DAE set model is expressed as,     
\begin{equation}
\label{StateFull}
\frac{dT}{dt}=g(T,v_\textrm{m})=\frac{1}{C_\textrm{th}}\cdot G_\textrm{eq}(T)\cdot v_\textrm{m}^2-\frac{T-T_\textrm{amb}}{R_\textrm{th}(T)C_\textrm{th}},
\end{equation}
where
\begin{equation}
\label{NewMemd}
G_\textrm{eq}(T)=\frac{G(T)}{(1+G(T)R_\textrm{e})^2},
\end{equation}
and
\begin{equation}
\label{imFull}
i_\textrm{m}=i=\frac{G(T)}{1+G(T)R_\textrm{e}}\cdot v_\textrm{m}=G_\textrm{m}(T)\cdot v_\textrm{m},
\end{equation}   
Based on \eqref{StateFull}, the input power $p$ dissipated on $M$, expressed as a function of $v_\textrm{m}$, and $G_\textrm{eq}(T)$, reads as,
\begin{equation}
\label{powerFull}
p=G_\textrm{eq}(T)\cdot v_\textrm{m}^2.
\end{equation}
Fig. \ref{Fig1}(c) illustrates the steady-state\footnote{The steady-state response represents the asymptotic behavior of a system, in our case study under a periodic voltage input, as time progresses toward infinity. A detailed analysis of both transient and steady-state responses of the $\textrm{NbO}_2$-Mott memristor, when excited by a high-frequency sinusoidal voltage input, is provided in sections \ref{sec4} and \ref{sec5}.} current-voltage characteristics obtained by applying a zero-mean sinusoidal voltage input $v_\textrm{i}$ to $M'$ with an amplitude of $\hat{v}_\textrm{i} = \SI{1.25}{V}$ and frequencies chosen from the set $f \in \{\SI{100}{KHz}, \SI{10}{MHz}, \SI{50}{MHz}, \SI{100}{MHz}\}$. Notably, the frequency dependence of the current-voltage plots in panel (c) seems to contradict a fundamental memristive fingerprint, according to which, the pinched hysteresis lobe area should decrease as the input frequency increases. In fact, as reported in \cite{9629238}, the $\textrm{NbO}_2$ current-voltage lobe area will keep increasing up to a critical frequency\footnote{Based on the theoretical analysis in \cite{9629238}, under the application of a sinusoidal voltage input with an amplitude of $\hat{v}_\textrm{m}=\SI{1.2}{V}$, the area of the current-voltage lobe of the memristor $M'$ reaches its maximum value when the input frequency becomes approximately equal to $1.2\cdot10^8\,\textrm{Hz}$. Beyond this critical value, the lobe area decreases abruptly.}, subsequently reducing towards the zero value as the input frequency tends to infinity. This is a characteristic behavior observed in memristors described by an energy balance equation similar to \eqref{NbOx} (or \eqref{StateFull}) (see \cite{Slesazeck2015}, \cite{6739164}). 

As a final note for this section, it's important to clarify that throughout the manuscript, whenever we refer to a voltage input applied to the nanodevice under investigation, it will imply a zero-mean sinusoidal voltage input, since the forthcoming behavioral observations and mathematical analyses will revolve around this type of stimulus.

\begin{figure}[t]
\centering
\includegraphics[width=0.4\linewidth]{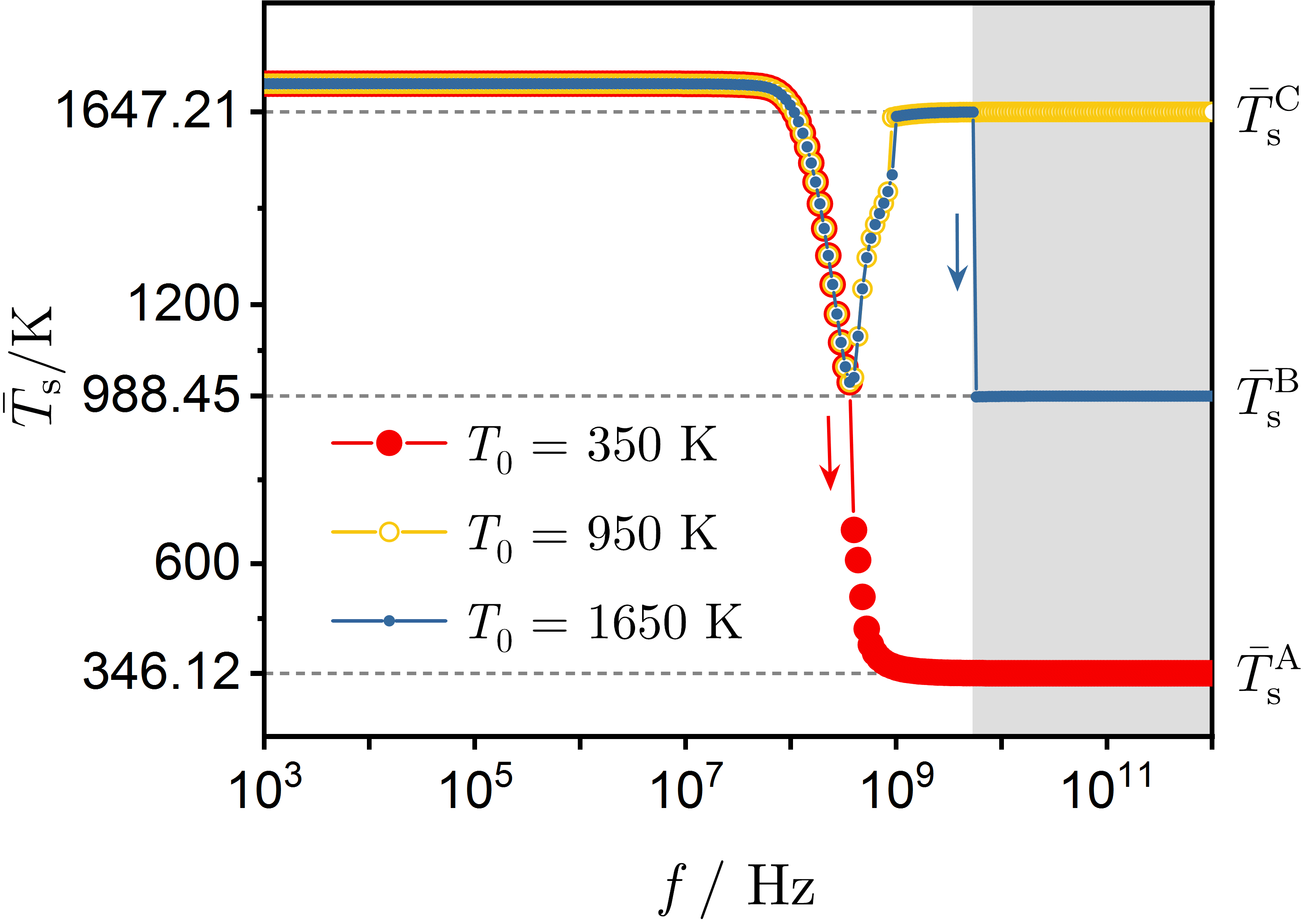}
\caption{Frequency induced bifurcation phenomenon in the $\textrm{NbO}_2$-Mott memristor under a zero-mean sinusoidal input of the form $v_i=v_\textrm{m}=(\SI{1.33}{V})\cdot\textrm{sin}(2\cdot\pi\cdot\ f\cdot t)$. The red, blue, and yellow-colored plots display the mean oscillating value $\bar{T}_\textrm{s}$ of the temperature time-waveform $T$, at steady-state, as a function of the input frequency $f$, within the range of $[10^3,10^{12}],\textrm{Hz}$, when the volatile memristor is initialized at temperatures $\SI{350}{K}$, $\SI{950}{K}$, and $\SI{1650}{K}$, respectively.}    
\label{Fig3}
\end{figure}

\section{Frequency-Induced Bifurcations}
\label{sec3}

The panels in Fig. \ref{Fig2} can be divided into rows, with the first, second, and third rows showing simulated steady-state current-voltage responses (see panels (a1), (b1), and (c1)) and steady-state temperature time-waveforms (see panels (a2), (b2), and (c2)) of the $\textrm{NbO}_2$-Mott memristor. These responses were obtained by applying periodic inputs across $M'$ (see Fig. \ref{Fig1}(b)) at frequencies of $10^3\,\textrm{Hz}$, $10^9\,\textrm{Hz}$, and $10^{11}\,\textrm{Hz}$, respectively, with an amplitude $\hat{v}_\textrm{i}$ defined as $\SI{1.33}{V}$. Additionally, each panel in Fig. \ref{Fig2} illustrates three curves representing different initial conditions $T_0$ for the device body temperature $T$. Specifically, for the red-, blue- and yellow-colored curves, $T_0$ was set to be equal to $\SI{350}{K}$, $\SI{950}{K}$, and $\SI{1650}{K}$. Notice from panels (a1) - (a2), that the nanodevice exhibits a single globally asymptotically stable steady-state behavior an utilized input frequency of $\SI{1}{KHz}$. As shown in (a2), the memristor temperature $T$ will oscillate periodically at steady-state around a mean value $\bar{T}_\textrm{s}$ (as indicated by the dashed horizontal line), regardless of the initial conditions. In fact, it can be easily verified that for this input frequency, the device will eventually settle to the same steady-state behavior regardless of the chosen value for $T_0$.
On the other hand, as shown in panels (b1) - (b2), for $f=10^9\,\textrm{Hz}$, the volatile memristor exhibits two distinct behaviors. Notably, the steady-state responses associated with the initial conditions $T_0=\SI{950}{K}$ and $T_0=\SI{1650}{K}$ are still merged as in (a1) - (a2). As the input frequency further increases, it displays three locally stable steady-states, as visualized by the three current-voltage plots in panel (c1), and by the three temperature time-responses in (c2). This frequency induced bifurcation phenomenon emerging in the $\textrm{NbO}_2$-Mott memristor for the selected initial conditions and type of input voltage waveform, is more clearly illustrated in Fig. \ref{Fig3}, which plots the mean oscillating value $\bar{T}_\textrm{s}$ of the temperature time-waveform $T$, at steady-state, as a function of the input frequency $f$, spanning the range $[10^3,10^{12}]\,\textrm{Hz}$. The first bifurcation occurs around $\SI{382.25}{MHz}$, marked by the separation of the red points from the blue and yellow ones, and indicated by the red arrow. The second bifurcation, resulting in three locally asymptotic stable steady-state behaviors, happens at approximately $\SI{5.2}{GHz}$, when the blue points deviate abruptly from the yellow ones as the input frequency increases (see the blue-colored arrow). \par
In the following sections, we will study the high-frequency response of the $\textrm{NbO}_2$-Mott memristor, focusing on the frequency range, where the device showcases three stable periodic behaviors, given the input characteristics and initial conditions assumed in this section. This behavior is exhibited for $f>\SI{5.2}{GHz}$ (see the gray-colored region in Fig. \ref{Fig3}).

\begin{figure*}[t]
\centering
\includegraphics[width=0.9\linewidth]{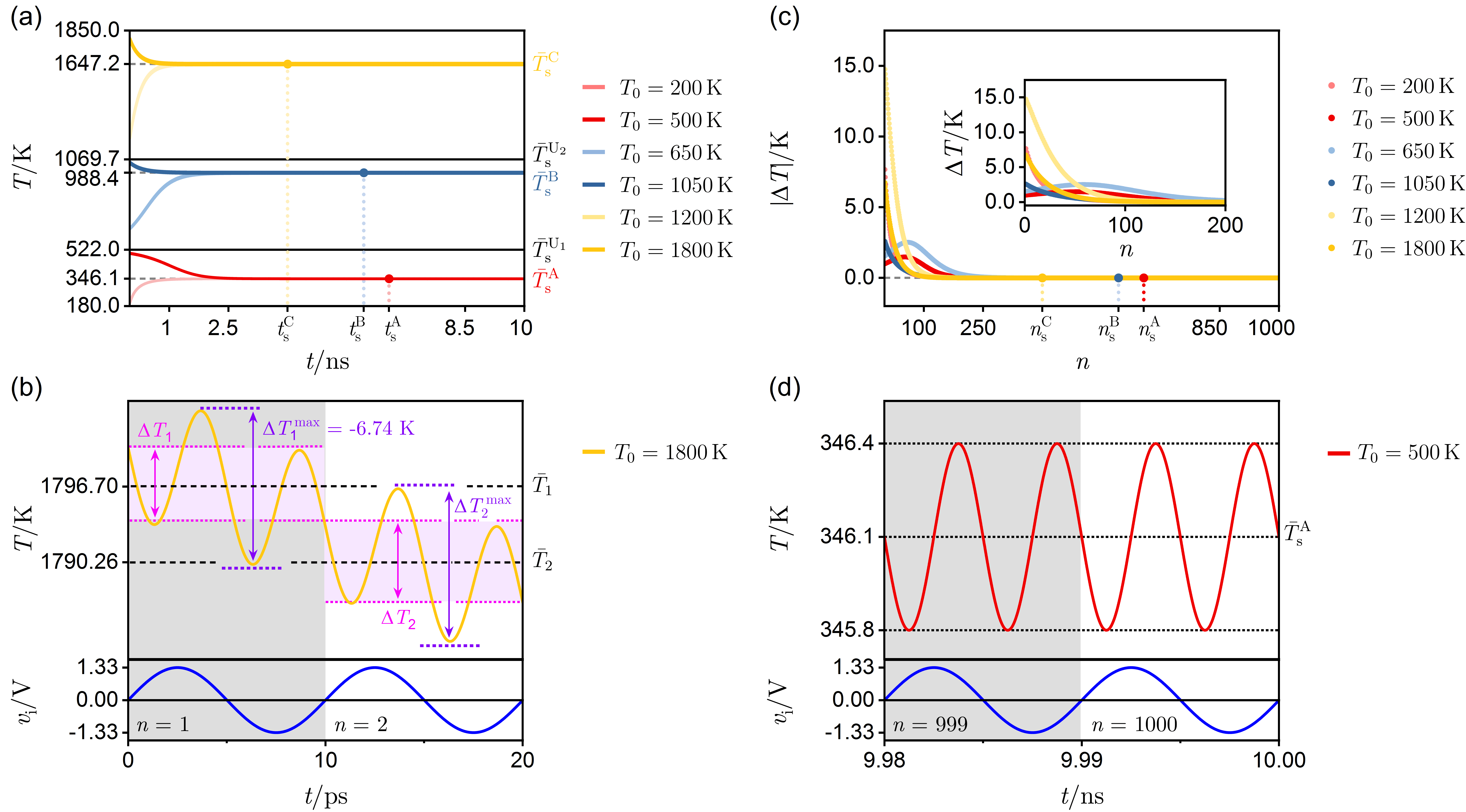}
\caption{High-frequency response of the $\textrm{NbO}_2$-Mott memristor. Panel (a): Time-series of the device body temperature $T$ under a sinusoidal voltage stimulus $v_\textrm{i}$ oscillating with an amplitude of $\hat{v}=\SI{1.33}{V}$ and a frequency of $f=\SI{100}{GHz}$, for different initial conditions $T_0$ (see the legend). The temperature levels $\bar{T}^\textrm{A}_\textrm{s}$, $\bar{T}^\textrm{B}_\textrm{s}$, and $\bar{T}^\textrm{C}_\textrm{s}$, depicted by the lower, middle, and upper horizontal dashed lines, represent the mean values of the red, blue, and yellow-shaded periodic temperature time-responses, at steady-state. The horizontal black lines at $\bar{T}^{\textrm{U}_1}_\textrm{s}$ and $\bar{T}^{\textrm{U}_2}_\textrm{s}$ separate the trajectories that lead to different asymptotic steady-state behaviors (for further details, refer to the analysis located towards the end of section \ref{sec3}). Symbols $t_\textrm{s}^\textrm{A}$, $t_\textrm{s}^\textrm{B}$, and $t_\textrm{s}^\textrm{C}$, represent the time-points where the red, blue, and yellow-shaded time-responses converge to the same periodic oscillation about $\bar{T}^\textrm{A}_\textrm{s}$, $\bar{T}^\textrm{B}_\textrm{s}$, and $\bar{T}^\textrm{C}_\textrm{s}$, respectively. Panel (b): Yellow-colored locus: Visualization and analysis of the $T$ versus $t$ plot, for $T_0=\SI{1800}{K}$, shown in (a), over the first two input cycles (see the blue-colored sinusoidal input voltage time-waveform plotted in the same panel). Symbol $\Delta T_1$ ($\Delta T_2$) indicates the net change of $T$ over the first (second) input cycle. Symbol $\Delta T_1^{\textrm{max}}$ ($\Delta T_2^{\textrm{max}}$) represents the maximum swing of $T$ over the first (second) input cycle, i.e. for $n=1$ ($n=2$). The temperature level $\bar{T}_1$ ($\bar{T}_2$), defined by the horizontal dashed line, is the mean oscillating value of $T$ over the first (second) input cycle. Panel (c): Net change $\Delta T$ of the device temperature $T$ over any input cycle $n$, for each of the temperature time-responses illustrated in panel (a). Symbols $n_\textrm{s}^\textrm{A}$, $n_\textrm{s}^\textrm{B}$, and $n_\textrm{s}^\textrm{C}$, represent the input cycle numbers where the red, blue, and yellow-shaded time-responses become periodic. The inset figure provides a detailed view of the $\Delta T$ values over the first 200 input cycles. Panel (d): Red-colored locus: Zoomed-in view of the time-response in (a), for $T_0=\SI{500}{K}$, over the last two input cycles (see the blue colored sinusoidal input voltage time-waveform plotted in the same panel).          
}    
\label{Fig4}
\end{figure*}

\section{High Frequency Response}
\label{sec4}
The time-responses plotted in red, blue, and yellow (light-red, light-blue, and light-yellow) colors in Fig. \ref{Fig4}(a) were obtained by integrating \eqref{StateFull}, assuming the application of a sinusoidal voltage input $v_\textrm{i}$ with an amplitude of $\hat{v}_\textrm{i}=\SI{1.33}{V}$ and a frequency of $f=\SI{100}{GHz}$ across $M'$ (see Fig. \ref{Fig1}(b)). The device body temperature was initialized at $T_0=\SI{500}{K}$, $\SI{1050}{K}$, and $\SI{1800}{K}$ ($\SI{200}{K}$, $\SI{650}{K}$, and $\SI{1200}{K}$) for the respective cases. Each simulation was performed over a total of $n=1000$ cycles. The yellow-colored time-series in Fig. \ref{Fig4}(b) zooms in over the first two periods of the $T$ versus $t$ plot for $T_0=\SI{1800}{K}$, shown in (a). As it may inferred from the illustrated time-response, the mean temperature values $\bar{T}_1$ and $\bar{T}_2$ (see the horizontal dashed lines) over the two depicted input cycles $n=1$ and $n=2$ are much larger than $\Delta T_1^\textrm{max}$ and $\Delta T_2^\textrm{max}$, denoting the maximum swing of $T$ over their respective cycles. Based on this behavioral observation, we may therefore assume that during the $n^\textrm{th}$ input cycle, the device temperature $T$ is approximately equal to the mean value $\bar{T}_n$ of its oscillation over the same cycle ($n\in N>0$). Let us now assume that $\Delta T_n$ is the net change of $T$ over the $n^\textrm{th}$ input cycle (see $\Delta T_1$ and $\Delta T_2$ in Fig. \ref{Fig4}(b)). Next, for the six temperature time-series depicted in Fig. \ref{Fig4}(a), we calculated the net temperature change $\Delta T_n$ over each input cycle $n$. Subsequently, we plotted each of these six sets of $(n, |\Delta T_n|)$ data pairs in Fig. \ref{Fig4}(c), with each data set retaining the same color as its corresponding curve in Fig. \ref{Fig4}(a). Notice that the illustrated plots, whether they decrease monotonically, or they peak before decaying, they all tend to approach the zero value asymptotically, as $n$ increases. In our study, we define the system as having reached steady-state when $\Delta T$ has dropped below $10^{-6}\,\textrm{K}$. As discussed in the previous section, the $T$ versus $t$ loci, illustrated in Fig. \ref{Fig4}(a), reveal an interesting feature of the $\textrm{NbO}_2$-Mott memristor, that is its capability to exhibit three locally-stable steady-state responses, under a high-frequency periodic voltage input, depending on the initial condition $T_0$\footnote{This phenomenon, referred to as local fading memory, which is present in multistable devices such as the $\textrm{NbO}_2$-Mott memristor, has been thoroughly investigated in \cite{7557034} and \cite{Ascoli2016}.}. The red-shaded time-waveforms converge asymptotically at time point $t_\textrm{s}^\textrm{A}$ to the same periodic oscillation around $\bar{T}^\textrm{A}_\textrm{s}$ (see Fig. \ref{Fig4}(d)). Similarly, the blue and yellow-shaded time-waveforms converge at time points $t_\textrm{s}^\textrm{B}$ and $t_\textrm{s}^\textrm{C}$ to common periodic oscillations around $\bar{T}^\textrm{B}_\textrm{s}$ and $\bar{T}^\textrm{C}_\textrm{s}$, respectively. The input cycle numbers $n_\textrm{s}^\textrm{A}$, $n_\textrm{s}^\textrm{B}$, and $n_\textrm{s}^\textrm{C}$, corresponding to time-points $t_\textrm{s}^\textrm{A}$, $t_\textrm{s}^\textrm{B}$, and $t_\textrm{s}^\textrm{C}$, are highlighted in Fig. \ref{Fig4}(c). The basins of attraction, within which the multistable $\textrm{NbO}_2$-Mott memristor admits the visualized steady-state trajectories, will be defined towards the end of this section.\par
\begin{figure}[t]
\centering
\includegraphics[width=0.35\linewidth]{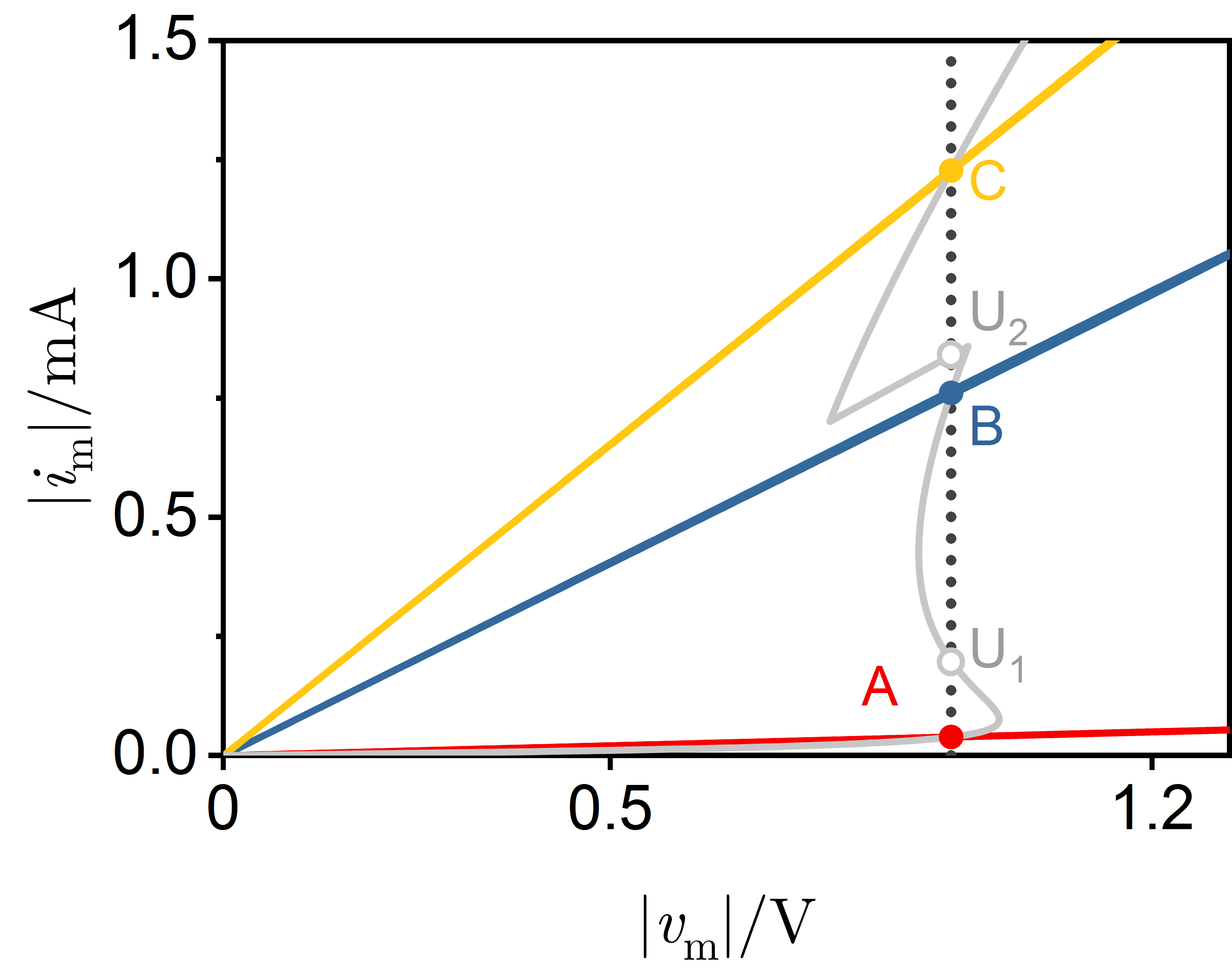}
\caption{Steady-state current-voltage responses of the $\textrm{NbO}_2$-Mott nanodevice, under the same sinusoidal voltage input and initial conditions as the ones utilized in Fig. \ref{Fig4} (see the legend). The gray-colored curve corresponds to the DC $I-V$ plot of the device. The dashed vertical line pinpoints the RMS voltage level  $V_\textrm{rms}$ of the employed sinusoidal voltage input, which is equal to $\SI{0.94}{V}$. Points A, B, and C, depict the intersection points of the red, blue, and yellow-colored current-voltage loci, respectively, with the DC $I_\textrm{m}-V_\textrm{m}$ plot, and the dashed vertical line. The DC $I_\textrm{m}-V_\textrm{m}$ plot, and the dashed vertical line, also intersect at points $\textrm{U}_1$ and $\textrm{U}_2$.}    
\label{Fig5}
\end{figure}

           
The red-colored time-series in Fig. \ref{Fig4}(d) zooms in across the last two periods of the $T$ versus $t$ plot for $T_0=\SI{500}{K}$, shown in panel (a). We notice, that the device body temperature responds to the applied periodic voltage input by oscillating periodically about a mean value $\bar{T}^\textrm{A}_\textrm{s}$ with a negligible peak-to-peak amplitude ($\approx \SI{0.6}{K}$), further verifying our assumption, this time during the steady-state phase of the temperature oscillation. 
As a matter of fact, even though our hypothesis was deduced from a simulation experiment that involved a high-frequency periodic voltage input of sinusoidal shape, it is valid, irrespective of the input voltage waveform shape and amplitude, provided that its frequency is high enough to endow the device with a single-valued current-voltage locus, over each input cycle, such as the ones shown in Fig. \ref{Fig5} \cite{9956789}. 
Specifically, the colored plots in this figure illustrate the three steady-state $|i_\textrm{m}|-|v_\textrm{m}|$ loci corresponding to the time-responses in Fig. \ref{Fig4}(a). We notice that the high value of the input frequency, gives rise to non-hysteretic device responses on the current-voltage plane. This behavior is a typical memristive fingerprint, as reported in \cite{Adhikari2013}. 
But why is it necessary to employ such a high input frequency ($f=\SI{100}{GHz}$) to observe the aforementioned distinctive memristive characteristics in this particular nanodevice? The main reason lies in its extremely small thermal capacitance ($C_\textrm{th}=10^{-16}\,\textrm{W}\cdot \textrm{s}\cdot \textrm{K}^{-1}$). This point will be further discussed over the course of the analysis performed in this section. \par
Considering the above-described behavioral observations, the net change $\Delta T$ in the device temperature $T$, over each cycle of a high-frequency periodic voltage input $v_\textrm{i}$, can be expressed as a function of the running mean value $\bar{T}$, over the same cycle, through the following integral:
\begin{equation}
\label{DeltaT}
\Delta T =\int_{0}^{1/f} g(\bar{T},v_\textrm{i}) \,dt=\frac{1}{f\cdot C_\textrm{th}}\left(\bar{p}-\frac{\bar{T}-T_\textrm{amb}}{R_\textrm{th}(\bar{T})}\right),
\end{equation}
where $g(\cdot)$ is the state evolution function (see \eqref{StateFull}), and  
\begin{equation}
\label{power1}
\bar{p}=f\cdot\int_{0}^{1/f} G_\textrm{eq}(\bar{T})\cdot v_\textrm{i}^2 \,dt,
\end{equation}
is the time-averaged power, delivered to a resistance equal to $1/G_\textrm{eq}(\bar{T})$, over each cycle by the high-frequency periodic voltage input $v_\textrm{i}$. It happens, that $\bar{p}$ can be calculated in terms of the root mean square voltage $V_\textrm{rms}$ of $v_\textrm{i}$ according to the following equation: 
\begin{equation}
\label{power2}
\bar{p}=G_\textrm{eq}(\bar{T})\cdot V_\textrm{rms}^2,
\end{equation}
since $G_\textrm{eq}(\bar{T})$ is constant over each input period. At this point, it is important to note that the calculation of $\bar{p}$ in terms of the effective input voltage $V_\textrm{rms}$ was possible due to the fact that the $\textrm{NbO}_2$-Mott nanodevice is a \textit{generic} memristor. Thus, its memductance function $G(\cdot)$ (and thus $G_\textrm{eq}(\cdot)$ through \eqref{NewMemd}), being independent of the input voltage $v_\textrm{i}$ (see \eqref{im}), remains constant and equal to $G(\bar{T})$ over each input cycle. Summarizing our mathematical investigation, by combining equations \eqref{DeltaT} and \eqref{power2}, $\Delta T$ reads as
\begin{equation}
\label{DeltaTfinal}
\Delta T =\frac{1}{f\cdot C_\textrm{th}}\cdot\left(G_\textrm{eq}(\bar{T})\cdot V_\textrm{rms}^2-\frac{\bar{T}-T_\textrm{amb}}{R_\textrm{th}(\bar{T})}\right),
\end{equation}
or
\begin{equation}
\label{DeltaTfinalEn}
\Delta T =\frac{1}{C_\textrm{th}}\cdot\Delta E,
\end{equation}
where 
\begin{equation}
\label{DeltaEn}
\Delta E=E_\textrm{el}-E_\textrm{th}.
\end{equation}
Parameter $E_\textrm{el}$ is the mean electrical energy provided to a resistance equal to $1/G_\textrm{eq}(\bar{T})$ by the periodic input signal over each cycle, and expressed as 
\begin{equation}
\label{ElectricalEnergy}
E_\textrm{el}=\frac{1}{f}\cdot{G_\textrm{eq}(\bar{T})\cdot V_\textrm{rms}^2},
\end{equation}
and $E_\textrm{th}$ defines the exchange of thermal energy with the surrounding environment over each input cycle, which reads as
\begin{equation}
\label{ThermalEnergy}
E_\textrm{th}=\frac{1}{f}\cdot\frac{\bar{T}-T_\textrm{amb}}{R_\textrm{th}(\bar{T})}.
\end{equation}
Essentially, \eqref{DeltaTfinal} - \eqref{DeltaEn} describe a typical electrothermal device behavior. In this context, the device temperature increases by $\Delta T$ during an input cycle when it receives more energy $E_\textrm{el}$ from an external power source than it loses as it releases thermal energy $E_\textrm{th}$ to the ambient environment, over the same cycle. Conversely, the temperature decreases when the energy loss due to thermal transfer to the environment exceeds the energy absorbed from the external source.
Since, the $\textrm{NbO}_2$-Mott nanodevice is a passive memristor, $E_\textrm{el}$ will be positive over each input cycle. On the other hand, the polarity of the thermal energy transferred to the ambient depends on the sign of $E_\textrm{th}$. 
Based on \eqref{ThermalEnergy}, for each input cycle, where $\bar{T}$ exceeds (is below) $T_\textrm{amb}$, the mean thermal energy $E_\textrm{th}$ is positive (negative), i.e. thermal energy is released to (is absorbed from) the ambient. 
The net change $\Delta T$ of the device temperature $T$ over each input cycle, where $E_\textrm{el}>E_\textrm{th}$ ($E_\textrm{el}<E_\textrm{th}$), is positive (negative). Based on these physical processes, irrespective of the initial device temperature $T_0$, over time, $E_\textrm{el}$ and $E_\textrm{th}$ become progressively closer one to the other with each input cycle, being practically balanced at $t_\textrm{s}$, where $\Delta E=E_\textrm{el}-E_\textrm{th}\approx 0$, and thus $\Delta T\approx 0$. At this time-point we assume that the temperature time-response has reached its steady-state. \par
As mentioned earlier in this section, the volatile memristor under study is characterized by an extremely small thermal capacitance value ($C_\textrm{th}=10^{-16}\,\textrm{W}\cdot \textrm{s}\cdot \textrm{K}^{-1}$). Physically, this means that any lack (excess) of thermal energy is swiftly balanced as the device absorbs it through Joule self-heating effects (as the device releases it to the ambient), leading to a simultaneous rapid increase (decrease) in the device body temperature $T$. If the input frequency is not high enough (roughly over $\SI{1}{GHz}$ in our case study), the temperature fluctuation during each cycle becomes non-negligible, and the assumptions used to derive \eqref{DeltaT} - \eqref{DeltaTfinal} are no longer valid. As a matter of fact, from \eqref{DeltaTfinal}, we notice that $\Delta T$ is inversely proportional to the product between the input frequency $f$ and the thermal capacitance $C_\textrm{th}$, i.e. the exceptionally small value of $C_\textrm{th}$ necessitates a correspondingly high value for $f$ to ensure that $\Delta T$ remains sufficiently small. 
Let us now investigate the predictive power of \eqref{DeltaTfinal} - \eqref{ThermalEnergy} and assess the extent to which these equations may be employed to characterize the response of the $\textrm{NbO}_2$-Mott memristor in the high-frequency limit.\par
\begin{figure}[t]
\centering
\includegraphics[width=0.55\linewidth]{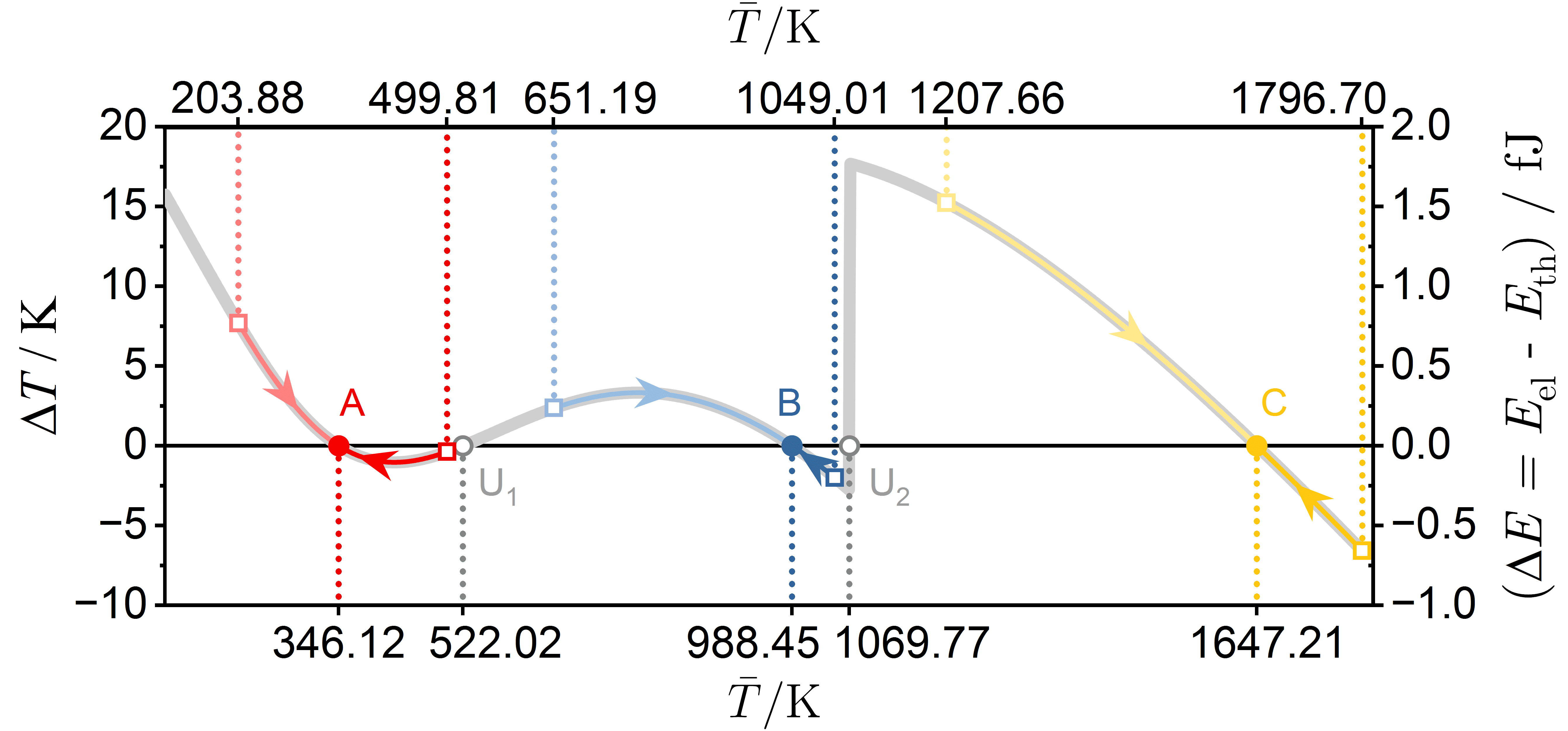}
\caption{The gray-colored $\Delta T$ ($\Delta E$) versus $\bar{T}\in[100,1800]\,\textrm{K}$ locus, whose ordinate values may be read along the left (right) vertical axis, was obtained through \eqref{DeltaTfinal} (\eqref{DeltaEn} - \eqref{ThermalEnergy}), for $V_\textrm{rms}=\SI{0.94}{V}$ and $f=\SI{100}{GHz}$. Colored (white-filled gray-shaded) round symbols A, B, and C ($\textrm{U}_1$ and $\textrm{U}_2$) depict the points where the gray-colored curve crosses the horizontal axis with negative (positive) slopes. Points $\textrm{U}_1$, and $\textrm{U}_2$ define the basins of attraction, within which the multistable $\textrm{NbO}_2$-Mott nanodevice exhibits three different steady-state responses, as visualized in Fig. \ref{Fig4}(a) (refer to the body text for more details). The red, blue, and yellow-shaded arrowed loci, found to lie over part of the gray-colored curve, were obtained through the simulations illustrated in Fig. \ref{Fig4}(a). The arrows indicate the transient evolution of the trajectory point $(\bar{T}, \Delta T)$ over 1000 input cycles for the simulated time-responses in Fig. \ref{Fig4}(a). Square-shaped white-filled symbols define the mean value $\bar{T}_1$ and the corresponding net change $\Delta T_1$ for each of the simulated time-responses in Fig. \ref{Fig4}(a).}    
\label{Fig6}
\end{figure}

  
The gray-shaded $\Delta T$ versus $\bar{T}\in[100,1800]\,\textrm{K}$ in Fig. \ref{Fig6} (refer to the left vertical axis to read the values for $\Delta T$) were obtained through \eqref{DeltaTfinal}, assuming that $V_\textrm{rms}=\hat{v}_\textrm{i}/\sqrt{2}=\SI{0.94}{V}$ ($\hat{v}_\textrm{i}=\SI{1.33}{V}$), and $f=\SI{100}{GHz}$. Based on the works in \cite{9956789,Pershin2019,Ascoli2023}, this plot represents the TA-SDR of the $\textrm{NbO}_2$-Mott memristor for a sinusoidal input characterized by an amplitude and frequency equal to $\hat{v}_\textrm{i}=\SI{1.33}{V}$ and $f=\SI{100}{GHz}$, respectively. The aforementioned literature sources provide detailed mathematical analyses, demonstrating how the input-referred TA-SDR can predict the time evolution of the mean value $\bar{T}$ of the state variable $T$ from any initial condition over a specified number of cycles of the high-frequency periodic input\footnote{The TA-SDR visualization tool, named in Ascoli \textit{et al.} \cite{Ascoli2023} and initially introduced by Pershin and Slipko in \cite{Pershin2019}, though not explicitly defined in the latter source, may be used to analyze the response of any first-order non-volatile memristor to square-wave periodic stimuli, which induce negligible changes in its state variable over each input cycle. In the meantime, Messaris \textit{et al.} \cite{9956789} presented a similar graphical tool, the high-frequency state dynamic route (HF-SDR), which generalized the methodology from \cite{Pershin2019} to characterize the response of any first-order non-volatile memristor to high-frequency periodic inputs of arbitrary forms. To avoid confusion, in this manuscript, we will adopt the nomenclature defined in \cite{Ascoli2023}.}. Notice, that the TA-SDR of Fig. \ref{Fig6} is referred to the same stimulus as the one employed for the simulations illustrated in Fig. \ref{Fig4}. As discussed earlier in this section (see \eqref{DeltaTfinal} - \eqref{ThermalEnergy}), the change in the device body temperature $\Delta T$ during each input cycle is determined by the net energy balance $\Delta E$, defined as the difference between the energy delivered to the device by the periodic power source $E_\textrm{el}$ and the thermal energy exchanged between the device itself and the ambient environment $E_\textrm{th}$, over the same cycle. According to \eqref{DeltaEn}, $\Delta T$ is equal to $\Delta E$ to within a scaling factor inversely proportional to the thermal capacitance $C_\textrm{th}$, allowing us to add a secondary vertical axis on the right hand side of the graph in Fig. \ref{Fig6}, to show $\Delta E=E_\textrm{el}-E_\textrm{th}$ versus $\bar{T}$. Thus, the TA-SDRs of electrothermal volatile memristor devices, such as the $\textrm{NbO}_2$-Mott threshold switch, may also offer insights into how these devices respond to the electrical energy generated by periodic voltage inputs, taking into account thermal gains and/or losses besides the device initial conditions. To facilitate the comparison of the calculations obtained through \eqref{DeltaTfinal} with simulation results, we extracted both the mean temperature value $\bar{T}$ and the net change in temperature $\Delta T$ over each input cycle for all six temperature time-series shown in Fig. \ref{Fig4}(a). Each of the six elicited data-sets of $(\bar{T},\Delta T)$ pairs was then plotted in Fig. \ref{Fig6}, while retaining the same color as its corresponding curve in panel (a) of Fig. \ref{Fig4}. Notably, the colored loci in Fig. \ref{Fig6} almost perfectly overlap the gray-shaded curve, verifying the accuracy of \eqref{DeltaTfinal} as well as the assumptions underlying its derivation. In other words, the gray-colored locus, plotted in Fig. \ref{Fig6}, predicts how the change $\Delta T$ in the device temperature per input cycle (refer to the left vertical axis to red its values) depends on the mean value $\bar{T}$ of the time waveform of the memristor temperature itself per input cycle, for any initial condition $T_0$, and for all cycles of the $\SI{100}{GHz}$ sine wave voltage stimulus from Fig. \ref{Fig4}(b). The arrow along each of the colored curves in Fig. \ref{Fig6} indicates the direction of the transient evolution of the corresponding point trajectory $(\bar{T},\Delta T)$ as it approaches asymptotically the temperature axis, where $\Delta T=0$, or equivalently, $\Delta E=0$. 
\begin{figure*}[t]
\centering
\includegraphics[width=1\linewidth]{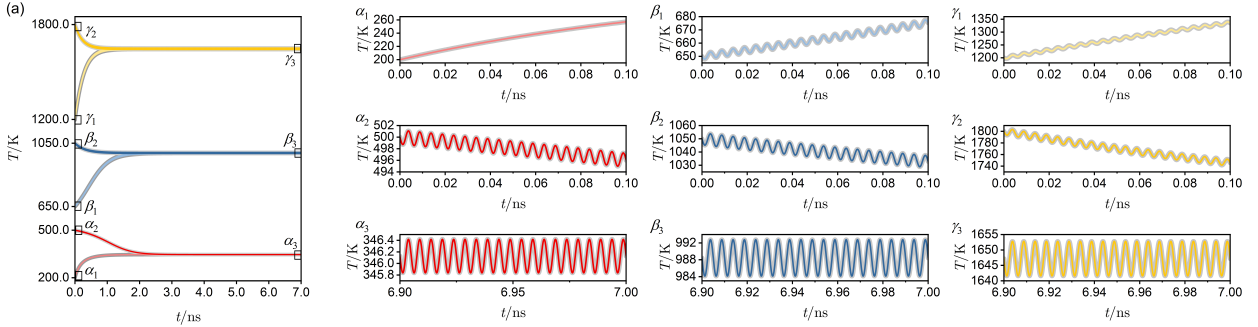}
\caption{Panel (a) compares simulated (see gray-colored waveforms) versus analytically calculated (see colored waveforms) temperature time-responses of the $\textrm{NbO}_2$-Mott memristor under a sinusoidal voltage input with an effective RMS voltage of $\SI{0.94}{V}$ oscillating at a frequency of $\SI{100}{GHz}$. The initial conditions are highlighted on the $y$-axis. Panels $\alpha_1$, $\beta_1$, and $\gamma_1$ ($\alpha_2$, $\beta_2$, and $\gamma_2$) provide a detailed view of the time responses in (a) initialized at a temperature $T_0$ equal to $\SI{200}{K}$, $\SI{650}{K}$, and $\SI{1200}{K}$ ($\SI{500}{K}$, $\SI{1050}{K}$, and $\SI{1800}{K}$), respectively, over the first 10 input cycles. Panels $\alpha_3$, $\beta_3$, and $\gamma_3$ zoom-in across the time responses in (a) oscillating periodically about the mean temperatures $\bar{T}_\textrm{s}^\textrm{A}=\SI{346.12}{K}$, $\bar{T}_\textrm{s}^\textrm{B}=\SI{988.45}{K}$, and $\bar{T}_\textrm{s}^\textrm{C}=\SI{1647.21}{K}$, respectively, over the last ten input cycles.}    
\label{Fig7}
\end{figure*}
It follows that the condition describing the periodic steady-state oscillation of $T$ can be derived by setting $\Delta T=0$ and substituting $\bar{T}$ with $\bar{T}_\textrm{s}$ in \eqref{DeltaTfinal}. The resulting relationship is expressed as
 \begin{equation}
\label{DeltaTfinalSteady}
g(\bar{T}_\textrm{s},V_\textrm{rms})=G_\textrm{eq}(\bar{T}_\textrm{s})\cdot V_\textrm{rms}^2-\frac{\bar{T}_\textrm{s}-T_\textrm{amb}}{R_\textrm{th}(\bar{T}_\textrm{s})}=0,
\end{equation}
where $\bar{T}_\textrm{s}$ is the mean value of the device body temperature oscillation, at steady-state. As expected, by setting $V_\textrm{rms}$ in \eqref{DeltaTfinalSteady} equal to $\SI{0.94}{V}$ and solving for $\bar{T}_\textrm{s}$, we obtain five solutions within the body temperature range of interest, in accordance to the gray-shaded curve in Fig. \ref{Fig6}, which goes through the $x$-axis five times within the same temperature range (see points A, $\textrm{U}_1$, B, $\textrm{U}_2$, and C, along the temperature axis in Fig. \ref{Fig6}). Note that in Fig. \ref{Fig6}, since $\Delta T$ is positive (negative) on the upper (lower) half-plane, the arrows along the $\Delta T$ versus $\bar{T}$ loci converge towards points A, B, and C, while they diverge away from points $\textrm{U}_1$ and $\textrm{U}_2$. Therefore, $M'$ exhibits three locally stable steady-state responses for the assumed high-frequency sinusoidal input. Each of these responses has a well-defined basin of attraction, where a unique asymptotic behavior emerges irrespective of the initial condition. Hence, from the illustration in Fig. \ref{Fig6}, it can be inferred that if the mean value $\bar{T}_1$ of the device body temperature $T$ over the first cycle of the assumed sinusoidal voltage input $v_\textrm{i}$ is lower than $\bar{T}_\textrm{s}^{\textrm{U}_1}$ (exceeds $\bar{T}_\textrm{s}^{\textrm{U}_2}$) equation \eqref{DeltaTfinalSteady} predicts that, at steady-state, the device body temperature will oscillate with a negligible peak-to-peak amplitude about the mean value $\bar{T}_\textrm{s}^\textrm{A}=\SI{346.12}{K}$ ($\bar{T}_\textrm{s}^\textrm{C}=\SI{1647.21}{K}$). If $\bar{T}$ over the first cycle of the stimulus takes values from the open set $(\bar{T}_\textrm{s}^{\textrm{U}_1},\bar{T}_\textrm{s}^{\textrm{U}_2})$, $T$ will oscillate periodically, at steady-state, about the mean value $\bar{T}_\textrm{s}^\textrm{B}=\SI{988.45}{K}$. Notice, that the basins of attraction associated to the TA-SDR equilibria $\textrm{A}$ and $\textrm{B}$ ($\textrm{B}$ and $\textrm{C}$) in Fig. \ref{Fig6}, are separated by the boundary point $\textrm{U}_1$ ($\textrm{U}_2$). Accordingly, the black horizontal lines in Fig. \ref{Fig4}(a), which correspond to the temperature levels $\bar{T}_\textrm{s}^{\textrm{U}_1}$ and $\bar{T}_\textrm{s}^{\textrm{U}_2}$, represent the separatrices located within the phase space of Fig. \ref{Fig6} at points $\textrm{U}_1$ and $\textrm{U}_2$, respectively.         
Since, for $T=\bar{T}_\textrm{s}$ and $v_\textrm{i}=V_\textrm{rms}$ the state evolution function $dT/dt=g(\cdot)$ is nullified (see \eqref{DeltaTfinalSteady}), the temperature-voltage pairs $(\bar{T}_\textrm{s}^j,V_\textrm{rms})$, where $j:\{\textrm{A},\textrm{U}_1,\textrm{B},\textrm{U}_2,\textrm{C}\}$, correspond to stationary states of $M'$, i.e. the points A, $\textrm{U}_1$, B, $\textrm{U}_2$, and C in Fig. \ref{Fig6}, can be also defined on the current-voltage plane exactly where the dashed vertical line at $V=V_\textrm{rms}=\SI{0.94}{V}$ (see Fig. \ref{Fig5}) intersects the DC $I_\textrm{m}-V_\textrm{m}$ plot of $M'$ (see the gray-colored curve in Fig. \ref{Fig5}). As a matter of fact, points A, B, and C ($\textrm{U}_1$ and $\textrm{U}_2$) correspond to stable (unstable) stationary states of the $\textrm{NbO}_2$-Mott memristor when driven by a DC voltage input equal to $\SI{0.94}{V}$. The interested reader may refer to a comprehensive investigation of the $\textrm{NbO}_2$-Mott memristor, including the methodology used to derive its DC $I_\textrm{m}-V_\textrm{m}$ plot and the characterization of the stability of its stationary operating points under both current and voltage DC stimulation, in the study performed in \cite{9629238}. It follows that the red, blue, and yellow-colored current-voltage loci illustrated in Fig. \ref{Fig5}, corresponding to the lower, middle, and upper locally stable steady-state responses shown in Fig. \ref{Fig4}(a), will pass through the points A, B, and C, respectively (see Fig. \ref{Fig5}). In fact, one can determine the maximum number of admissible stable or unstable steady-state behaviors for a volatile electrothermal memristor, described by an energy balance equation similar to \eqref{NbOx} or \eqref{StateFull} (see \cite{Brown2022,Demirkol2022a,Hennen2018,9629238,Ascoli2021a,Slesazeck2015,Pickett2012}), by identifying a particular voltage interval along the horizontal axis, through each point of which one may draw a vertical line, which crosses the DC $I_\textrm{m}-V_\textrm{m}$ the largest number of time. In particular, the intersections points, corresponding to stable stationary states, define the maximum number of observable steady-state oscillations under a periodic voltage input of sufficiently-high frequency, irrespective if it is sinusoidal, triangular, rectangular etc, as long as its RMS value is selected from the earlier specified voltage interval.         

\section{Analytical Derivation of the Transient and Steady-State Response}
\label{sec5}

Interestingly, the assumptions made so far may allow for the derivation of a closed-form formula that can accurately approximate the time-waveform of the $\textrm{NbO}_2$-Mott memristor body temperature, starting from any initial condition, in response to a high-frequency sinusoidal voltage input. Let us demonstrate how this can be achieved. \par
Assuming that $T_{0,n}$ $(\bar{T}_n)$ is the initial (mean) temperature value at the start of (over) the $n^\textrm{th}$ cycle of the input $v_\textrm{i}$, 
by substituting $G_\textrm{eq}(T)$, $R_\textrm{th}(T)$ and $v_\textrm{m}$ in equation \eqref{StateFull}, with $G_\textrm{eq}(\bar{T}_n)$, $R_\textrm{th}(\bar{T}_n)$ and $v_\textrm{i}=\hat{v}_\textrm{i}\cdot\textrm{sin}(2\cdot\pi\cdot f\cdot t)$, respectively, and subsequently integrating $dT/dt$ over time, the time evolution $T_n(t)$ of the device body temperature $T$ during the $n^\textrm{th}$ cycle of the high-frequency periodic input, that is for $t\in[(n-1)/f, n/f]$, is expressed as     
\begin{equation}
\label{TvstAnalytical}
T_n(t) =\delta_n+\epsilon_n\cdot\textrm{sin}(4\cdot\pi\cdot f\cdot t+\phi_n)+\zeta_n\cdot\textrm{exp}\left(-\frac{t}{\tau_n}\right),
\end{equation}
where $n\in N>0$, while
\begin{equation}
\label{delta}
\delta_n=\bar{T}_n+\Delta{T}_n\cdot f \cdot\tau_n,
\end{equation}
\begin{equation}
\label{epsilon}
\epsilon_n=\frac{\delta_n-T_\textrm{amb}}{\sqrt{1+16\cdot\pi^2\cdot f^2\cdot\tau^2_n}},
\end{equation}
\begin{equation}
\label{zeta}
\zeta_n=\frac{T_{0,n}-T_\textrm{amb}-16\cdot\pi^2\cdot f^2\cdot\tau^2_n\cdot(\delta_n-T_{0,n})}{1+16\cdot\pi^2\cdot f^2\cdot\tau^2_n},
\end{equation}
and
\begin{equation}
\label{phi}
\phi_n=\textrm{arccos}\left(\frac{4\cdot\pi\cdot f\cdot\tau_n}{\sqrt{1+16\cdot\pi^2\cdot f^2\cdot\tau^2_n}}\right)+\pi.
\end{equation}
\begin{figure}[t]
\centering
\includegraphics[width=0.35\linewidth]{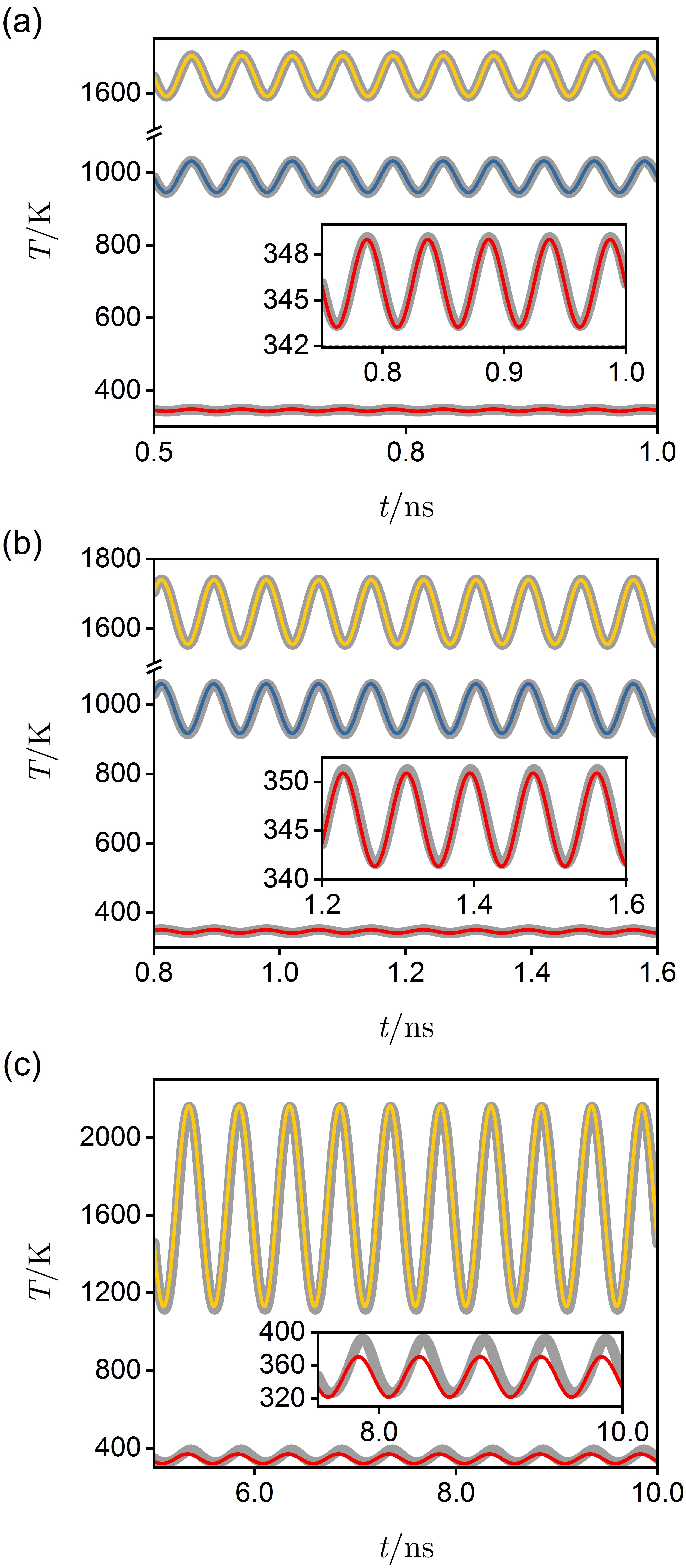}
\caption{Panels (a), (b), and (c) showcase steady-state temperature time-responses of the volatile memristor under study. The simulated responses are represented by gray-colored curves, while the analytically calculated ones are visualized by colored curves. These responses were obtained by applying a sinusoidal voltage input with an effective voltage $V_\textrm{rms}$ of $\SI{0.94}{V}$ across the device $M'$ (see Fig. \ref{Fig1}(b)). The input frequencies $f$ are $\SI{20}{GHz}$, $\SI{6}{GHz}$, and $\SI{1}{GHz}$ for panels (a), (b), and (c) respectively. For the displayed steady-state time-waveforms in the lower, middle, and upper sections of panels (a) and (b), the initial condition $T_0$ (parameter $\bar{T}_\textrm{s}$ in \eqref{TvstAnalyticalSS} - \eqref{phiSS}) for the gray (colored) curves is set to $\bar{T}^\textrm{A}_\textrm{s}$, $\bar{T}^\textrm{B}_\textrm{s}$, and $\bar{T}^\textrm{C}_\textrm{s}$, respectively. In panel (c), $T_0$ ($\bar{T}_\textrm{s}$) for the gray-colored (colored) time-responses in the lower and upper sections was set to $\bar{T}^\textrm{A}_\textrm{s}$ and $\bar{T}^\textrm{C}_\textrm{s}$, respectively. In each plot an inset provides a close-up view of the steady-state time course of the lower oscillation in the body temperature $T$.}    
\label{Fig8}
\end{figure}
\noindent The parameter $\Delta T_n$, which represents the net change in the device temperature $T$ over the $n^\textrm{th}$ cycle of the applied sinusoidal input $v_\textrm{i}$, may be calculated through \eqref{DeltaTfinal} after substituting $\bar{T}$ with $\bar{T}_n$. As mentioned at the beginning of this section, $T_{0,n}$ is the device temperature at the start of the $n^\textrm{th}$ input cycle, as defined via
\begin{equation}
\label{T0initial}
T_{0,n}=
\begin{cases} 
T_1(0)=T_0,\,\textrm{for}\,\,n=1\\
T_{n-1}\left(\frac{n-1}{f}\right),\,\textrm{for}\,\,n\geq2.\\
\end{cases}
\end{equation}
The parameter $\tau_n$ represents the device thermal time constant \cite{Brown2022b} during the $n^\textrm{th}$ input cycle, which is expressed as 
\begin{equation}
\label{tau}
\tau_n = R_\textrm{th}(\bar{T}_n) \cdot C_\textrm{th}.
\end{equation}
The thermal time constant $\tau_n$ regards to the time evolution of $T_n$ when the device is in the metallic (insulating) state if $\bar{T}_n > T_\textrm{C}$ $(\bar{T}_n \leq T_\textrm{C})$. \par
The time-evolution $T_1(t)$ of the device body temperature $T$ during the first cycle of the sinusoidal input for a predefined initial condition $T_0$ can be determined using the system of equations \eqref{TvstAnalytical} - \eqref{tau} for $n=1$ provided that the mean oscillating value over the first input cycle $\bar{T}_1$ is already known. To facilitate this process, we can assume $\bar{T}_1$ is equal to $T_0$, as simulation results indicate that these two values are very close to each other in the high-frequency limit. Notice, that $\bar{T}_1$ and $T_0$, for the simulation depicted in Fig. \ref{Fig4}(b), are equal to $\SI{1796.70}{K}$ and $\SI{1800}{K}$, respectively. An even more accurate $\bar{T}_1$ value can be obtained through the following equation:
\begin{equation}  
\label{Tmean}
\begin{split}
\bar{T}_1 & =f\cdot\int_{0}^{1/f} T_1(t) \,dt \\ 
&=\delta_1+f\cdot\epsilon_1\cdot\tau_1\cdot\Bigg(1-\textrm{exp}\left(-\frac{1}{\tau_1\cdot f}\right)\Bigg),
\end{split}
\end{equation}
where $\delta_1$, $\epsilon_1$, and $\tau_1$, are calculated through \eqref{delta}, \eqref{epsilon}, and \eqref{tau}, after setting $n=1$. Notice, that $\delta_1$, $\epsilon_1$, and $\tau_1$, depend on $\Delta T_1$, which, in turn, is linked to $\bar{T}_1$ through \eqref{DeltaTfinal} (refer to the text below \eqref{phi}). An approximate analytical solution for $\bar{T}_1$ may be obtained by substituting $\bar{T}_1$ with $T_0$ on the right hand side of \eqref{Tmean}. This is a reasonable approximation, based on the discussion performed a bit earlier in this section. Considering again the system simulated in Fig. \ref{Fig4}(b), $\bar{T}_1$ is calculated analytically in this manner to the value $\SI{1796.71}{K}$, which almost matches the one $\SI{1796.70}{K}$ obtained by numerically integrating the device DAE set model (see Fig. \ref{Fig4}(b)). Next, the mean value $\bar{T}_2$ is defined as the sum of $\bar{T}_1$ and $\Delta T_1$, i.e. $\bar{T}_2=\bar{T}_1+\Delta T_1$. Thus, the time-waveform $T_2(t)$ for the second cycle of the input stimulus can be derived through \eqref{TvstAnalytical} - \eqref{tau} with $n=2$. The same process can be repeated for any subsequent input cycle. \par
Fig. \ref{Fig7}(a) re-plots the simulated temperature time-responses shown in Fig. \ref{Fig4}(a), this time using a gray color for each waveform, instead of the previously employed colored visualization. On the other hand, each colored time-series, plotted in Fig. \ref{Fig7}(a), was generated by employing the just mentioned iterative procedure, utilizing the same initial condition $T_0$ and input voltage $v_\textrm{i}$, as the ones used for the corresponding gray-colored simulated waveform. Each colored time-response was obtained by utilizing \eqref{TvstAnalytical} - \eqref{tau} to reproduce $T_n(t)$ for each input cycle $n$. Panels $\alpha_1$, $\alpha_2$, $\alpha_3$, $\beta_1$, $\beta_2$, $\beta_3$, $\gamma_1$, $\gamma_2$, and $\gamma_3$, in Fig. \ref{Fig7}, present a zoom-in view of the temperature time-series featured in panel (a), enabling a detailed assessment of the simulated and analytically reproduced plots during both their transient and steady-state phases. The excellent accuracy of the proposed methodology is clearly demonstrated. Notice, that as the memristor evolves towards the steady-state phase, with each input period $n$, the exponential function on the right-hand side of \eqref{TvstAnalytical} progressively decays towards zero. At steady-state, we can assume its contribution to be negligible, practically zero, and thus the third term on the right-hand side of \eqref{TvstAnalytical} can be safely ignored. Therefore, the asymptotic oscillation $T_\textrm{s}(t)$ of the device body temperature $T$ about a mean value $\bar{T}_\textrm{s}$, characterized by a null net change $\Delta T_n$ in $T$ over each input steady-state cycle $n$, can be described by the following periodic function:  
\begin{equation}
\label{TvstAnalyticalSS}
T_\textrm{s}(t) =\delta_\textrm{s}+\epsilon_\textrm{s}\cdot\textrm{sin}(4\cdot\pi\cdot f\cdot t+\phi_\textrm{s}),
\end{equation}
where 
\begin{equation}
\label{deltaSS}
\delta_\textrm{s}=\bar{T}_\textrm{s},
\end{equation}
\begin{equation}
\label{epsilonSS}
\epsilon_\textrm{s}=\frac{\bar{T}_\textrm{s}-T_\textrm{amb}}{\sqrt{1+16\cdot\pi^2\cdot f^2\cdot\tau^2_\textrm{s}}},
\end{equation}
\begin{equation}
\label{phiSS}
\phi_\textrm{s}=\textrm{arccos}\left(\frac{4\cdot\pi\cdot f\cdot\tau_\textrm{s}}{\sqrt{1+16\cdot\pi^2\cdot f^2\cdot\tau^2_\textrm{s}}}\right)+\pi.
\end{equation}
Parameters $\epsilon_\textrm{s}$ and $\phi_\textrm{s}$ are the amplitude and phase of $T_\textrm{s}(t)$, while $\tau_\textrm{s} = R_\textrm{th}(\bar{T}_\textrm{s}) \cdot C_\textrm{th}$ is the thermal time constant at steady-state.\par
Up to this point, we have confirmed the accuracy of the proposed methodology and of the derived analytical equations when the input frequency is as high as $\SI{100}{GHz}$. Let us now determine the lower frequency limit for the validity of our assumptions, specifically focusing on the applicability of equations \eqref{TvstAnalyticalSS} - \eqref{phiSS}. Panels (a), (b), and (c) in Fig. \ref{Fig8} compare simulated (see gray-colored curves) and analytically calculated (see colored curves) temperature time-responses under a sinusoidal voltage input with an effective voltage $V_\textrm{rms}$ of $\SI{0.94}{V}$ at frequencies $f$ of $\SI{20}{GHz}$, $\SI{6}{GHz}$, and $\SI{1}{GHz}$, respectively. The initial condition $T_0$ (The parameter  $\bar{T}_\textrm{s}$ in \eqref{TvstAnalyticalSS} - \eqref{phiSS}) for the gray-colored (colored) steady-state time-waveform plotted in the lower, middle, and upper parts of panels (a) and (b) was set equal to $\bar{T}^\textrm{A}_\textrm{s}$, $\bar{T}^\textrm{B}_\textrm{s}$, and $\bar{T}^\textrm{C}_\textrm{s}$, respectively. In each case the selected $T_0$ value ensured that each of the numerically obtained time-response exhibits a periodic waveform for $t\geq0$. As anticipated, the accuracy of the proposed analytical equations reduces as the input frequency decreases, still remaining quite precise, though, for the whole frequency range where the $\textrm{NbO}_2$-Mott memristor exhibits three distinct steady-state behaviors, i.e. above $f=\SI{1}{GHz}$. As shown in Fig. \ref{Fig7}(c), at this frequency, the periodically-forced device exhibits two steady-state behaviors instead of the three observed for higher frequencies (this is also indicated in Fig. \ref{Fig3}). In particular, the initial device temperature $T_0$ (the mean oscillating value $\bar{T}_\textrm{s}$ of the device temperature $T$, at steady state) for the simulated gray-colored (analytically derived colored) time-responses seen in the lower and upper parts of Fig. \ref{Fig7}(c) was given the values $\bar{T}^\textrm{A}_\textrm{s}$ and $\bar{T}^\textrm{C}_\textrm{s}$, respectively. All in all, this section has demonstrated the accuracy of the proposed analytical model for the response of the $\textrm{NbO}_2$-Mott memristor to high-frequency sine wave stimuli.  
\section{Discussion and Conclusions}
\label{sec6}
Complementing our recent theoretical study on the high-frequency response of non-volatile memristors \cite{9956789}, this work explores the behavior of electro-thermal threshold switches, classified as volatile memristor devices, in the high-frequency limit. Although the methods detailed in this manuscript are applied to the HP $\textrm{NbO}_2$-Mott nano-device, they can be extended to all electrothermal devices, falling into the class of \textit{generic} memristors \cite{Chua2015}, and described by an energy balance equation similar to \eqref{NbOx} or \eqref{StateFull} (see \cite{Brown2022, Demirkol2022a, 9629238, 9181036,Ascoli2021a}). Unlike non-volatile memristors, the volatile $\textrm{NbO}_2$-Mott memristor is not programmable. In fact, despite, subjected to external stimuli, its resistance can be configured, it reverts to a default state once powered off. Even so, we showed that this threshold switch is quite versatile, displaying a maximum of three distinct asymptotic steady-state oscillatory behaviors when subjected to an appropriate high-frequency zero-mean periodic voltage input, with the specific oscillatory behavior it settles onto, being determined by its initial body temperature (see section \ref{sec3}). Quite counter-intuitively, as elucidated by the deep circuit-theoretic analysis performed in section \ref{sec4}, this capability is inscribed in its DC $I_\textrm{m}-V_\textrm{m}$ plot. Importantly, in the same section, it was demonstrated how a TA-SDR for the $\textrm{NbO}_2$-Mott memristor, and thus for any electrothermal threshold switch classified as a generic memristor, can be derived analytically (see \eqref{DeltaTfinal}) for any high-frequency periodic sine-wave voltage input. Let us stress at this point that the input-referred TA-SDR (see Fig. \ref{Fig6}) is particularly useful for the study of volatile threshold switches, being endowed with at least a couple of basins of attraction under a high-frequency periodic sine-wave voltage input. The input-referred TA-SDR depicts the locations of all the possible attractors in the one-dimensional memristor state space, around which the memristive system may oscillate at steady-state, and demarcates regions in the same state space, in each of which any initial condition leads to a different asymptotic oscillatory behavior for the device. By examining trajectories within these basins, one can gain insights into how the memristive system evolves toward its periodic steady states. On the other hand, the boundaries between the basins known as separatrices, correspond to admissible yet unstable periodic steady-states for the periodically-forced device. By linking the net change $\Delta T$ in the device temperature $T$ over an input cycle, with the mean value $\bar{T}$ of the temperature oscillation, over the same cycle, the input-referred TA-SDR illustrates how the high-frequency input stimulus maps each admissible temperature state of the $\textrm{NbO}_2$-Mott memristor from the beginning to the end of the cycle itself. Most importantly, the mathematical investigation presented in section \ref{sec4} indicates that this extremely useful visualization tool can be derived analytically on the basis of a closed-form formula (see \eqref{DeltaTfinal}), irrespective of the periodic voltage input type, be it sinusoidal, triangular, rectangular etc, as long as the input itself oscillates with a frequency that is high enough to induce negligible changes in the device body temperature over each input cycle. Going one step further, in Section \ref{sec5}, we introduced an algorithm, that reproduces the temperature time-waveform of the $\textrm{NbO}_2$-Mott nano-device, induced by a zero-mean sinusoidal voltage input oscillating in the high-frequency limit. 
We demonstrated that the steady-state time-waveform of the body temperature of the volatile memristor under study, subjected to a sufficiently-high frequency zero-mean sinusoidal voltage input, can be described by a closed-form analytical formula, which is rather accurate within the frequency range, where the $\textrm{NbO}_2$-Mott nano-device exhibits tristability.\par
Our recent study \cite{9956789} on the high-frequency response of non-volatile memristors, combined with the findings in this manuscript, have clearly demonstrated that memristors tend to operate as specified by their linearized models, when subjected to high-frequency AC periodic inputs. This discovery offers new opportunities for the application of methods from nonlinear system theory, potentially enhancing our comprehension of the device behavior, and enabling us to model its responses to arbitrary inputs from given classes more effectively. Consequently, by exhibiting a linearized behavior under high-frequency AC periodic inputs, memristors can be treated as weakly nonlinear systems, which enables their modelling with relatively low complexity.\par
Concluding, within the context of ongoing research, focused on pursuing robust predictive simulations of memristor-based circuits, the proposed analytical equations, capturing multistability in volatile threshold switches induced by high-frequency periodic inputs, could support the design of circuits, implementing innovative sensing and mem-computing paradigms, aimed at processing high-frequency oscillating signals for applications on the edge.

\bibliographystyle{unsrt}  
\bibliography{references}

\begin{thebibliography}{10}

\bibitem{Song2023}
Min-Kyu Song et~al.
\newblock {Recent Advances and Future Prospects for Memristive Materials, Devices, and Systems}.
\newblock {\em ACS Nano}, 17(13):11994--12039, jul 2023.

\bibitem{Brown2022}
Timothy~D. Brown, Stephanie~M. Bohaichuk, Mahnaz Islam, Suhas Kumar, Eric Pop, and R.~Stanley Williams.
\newblock {Electro-Thermal Characterization of Dynamical VO2 Memristors via Local Activity Modeling}.
\newblock {\em Advanced Materials}, 2022.

\bibitem{Demirkol2022a}
Ahmet~Samil Demirkol, Alon Ascoli, Ioannis Messaris, and Ronald Tetzlaff.
\newblock {Pattern formation dynamics in a Memristor Cellular Nonlinear Network structure with a numerically stable VO2 memristor model}, 2022.

\bibitem{Hennen2018}
T.~Hennen, D.~Bedau, J.~A.~J. Rupp, C.~Funck, S.~Menzel, M.~Grobis, R.~Waser, and D.~J. Wouters.
\newblock Forming-free mott-oxide threshold selector nanodevice showing s-type ndr with high endurance (> 1012 cycles), excellent vth stability (5
\newblock In {\em 2018 IEEE International Electron Devices Meeting (IEDM)}, pages 37.5.1--37.5.4, 2018.

\bibitem{9629238}
Ioannis Messaris, Timothy~D. Brown, Ahmet~S. Demirkol, Alon Ascoli, M.~Moner Al~Chawa, R.~Stanley Williams, Ronald Tetzlaff, and Leon~O. Chua.
\newblock Nbo2-mott memristor: A circuit- theoretic investigation.
\newblock {\em IEEE Transactions on Circuits and Systems I: Regular Papers}, 68(12):4979--4992, 2021.

\bibitem{Ascoli2021a}
Alon Ascoli, Ahmet~S. Demirkol, Ronald Tetzlaff, Stefan Slesazeck, Thomas Mikolajick, and Leon~O. Chua.
\newblock {On Local Activity and Edge of Chaos in a NaMLab Memristor}.
\newblock {\em Frontiers in Neuroscience}, 15, 2021.

\bibitem{Slesazeck2015}
S.~Slesazeck, H.~Mähne, H.~Wylezich, A.~Wachowiak, J.~Radhakrishnan, A.~Ascoli, R.~Tetzlaff, and T.~Mikolajick.
\newblock Physical model of threshold switching in nbo2 based memristors.
\newblock {\em RSC Adv.}, 5:102318--102322, 2015.

\bibitem{Pickett2012}
Matthew~D Pickett and R~Stanley Williams.
\newblock Sub-100 fj and sub-nanosecond thermally driven threshold switching in niobium oxide crosspoint nanodevices.
\newblock {\em Nanotechnology}, 23(21):215202, may 2012.

\bibitem{Kumar2020}
Suhas Kumar, R.~Stanley Williams, and Ziwen Wang.
\newblock {Third-order nanocircuit elements for neuromorphic engineering}.
\newblock {\em Nature}, 2020.

\bibitem{Pickett2013}
Matthew~D. Pickett, Gilberto Medeiros-Ribeiro, and R.~Stanley Williams.
\newblock {A scalable neuristor built with Mott memristors}.
\newblock {\em Nature Materials}, 2013.

\bibitem{Yi2018}
Wei Yi, Kenneth~K. Tsang, Stephen~K. Lam, Xiwei Bai, Jack~A. Crowell, and Elias~A. Flores.
\newblock {Biological plausibility and stochasticity in scalable VO 2 active memristor neurons}.
\newblock {\em Nature Communications}, 9(1), 2018.

\bibitem{Wang2018}
Zhongrui Wang, Rivu Midya, Saumil Joshi, Hao Jiang, Can Li, Peng Lin, Wenhao Song, Mingyi Rao, Yunning Li, Mark Barnell, Qing Wu, Qiangfei Xia, and J.~Joshua Yang.
\newblock Unconventional computing with diffusive memristors.
\newblock In {\em 2018 IEEE International Symposium on Circuits and Systems (ISCAS)}, pages 1--5, 2018.

\bibitem{9689062}
Alon Ascoli, Ahmet~Samil Demirkol, Ronald Tetzlaff, and Leon Chua.
\newblock Edge of chaos theory resolves smale paradox.
\newblock {\em IEEE Transactions on Circuits and Systems I: Regular Papers}, 69(3):1252--1265, 2022.

\bibitem{9855410}
Alon Ascoli, Ahmet~Samil Demirkol, Ronald Tetzlaff, and Leon Chua.
\newblock Edge of chaos is sine qua non for turing instability.
\newblock {\em IEEE Transactions on Circuits and Systems I: Regular Papers}, 69(11):4596--4609, 2022.

\bibitem{Ascoli2016Erase}
Alon Ascoli, Ronald Tetzlaff, Leon~O. Chua, John~Paul Strachan, and Richard~Stanley Williams.
\newblock {History Erase Effect in a Non-Volatile Memristor}.
\newblock {\em IEEE Transactions on Circuits and Systems I: Regular Papers}, 2016.

\bibitem{Yang2010}
J~Joshua Yang, M~X Zhang, John~Paul Strachan, Feng Miao, Matthew~D Pickett, Ronald~D Kelley, G~Medeiros-Ribeiro, and R~Stanley Williams.
\newblock {High switching endurance in TaOx memristive devices}.
\newblock {\em Applied Physics Letters}, 97(23), 2010.

\bibitem{Strachan2013}
John~Paul Strachan, Antonio~C. Torrezan, Feng Miao, Matthew~D. Pickett, J.~{Joshua Yang}, Wei Yi, Gilberto Medeiros-Ribeiro, and R.~{Stanley Williams}.
\newblock {State dynamics and modeling of tantalum oxide memristors}.
\newblock {\em IEEE Transactions on Electron Devices}, 2013.

\bibitem{9956789}
Ioannis Messaris, Alon Ascoli, Ahmet~S Demirkol, and Ronald Tetzlaff.
\newblock {High Frequency Response of Non-Volatile Memristors}.
\newblock {\em IEEE Transactions on Circuits and Systems I: Regular Papers}, 70(2):566--578, 2023.

\bibitem{Ascoli2018}
Alon Ascoli, Ronald Tetzlaff, and Stephan Menzel.
\newblock {Exploring the Dynamics of Real-World Memristors on the Basis of Circuit Theoretic Model Predictions}.
\newblock {\em IEEE Circuits and Systems Magazine}, 2018.

\bibitem{AscoliAdv}
Alon Ascoli, Stephan Menzel, Vikas Rana, Tim Kempen, Ioannis Messaris, Ahmet~Samil Demirkol, Michael Schulten, Anne Siemon, and Ronald Tetzlaff.
\newblock A deep study of resistance switching phenomena in taox reram cells: System-theoretic dynamic route map analysis and experimental verification.
\newblock {\em Advanced Electronic Materials}, 8(8):2200182, 2022.

\bibitem{Pershin2019}
Y.~V. Pershin and V.~A. Slipko.
\newblock {Bifurcation analysis of a TaO memristor model}.
\newblock {\em Journal of Physics D: Applied Physics}, 52(50), 2019.

\bibitem{Ascoli2023b}
A.~Ascoli, N.~Schmitt, I.~Messaris, A.~S. Demirkol, S.~Menzel, V.~Rana, R.~Tetzlaff, and L.~O. Chua.
\newblock System-theoretic analysis of bistability in the response of a tao x reram to pulse train stimuli.
\newblock {\em Frontiers in Nanotechnology}, 3, 2023.

\bibitem{Ascoli2023}
A.~Ascoli, N.~Schmitt, I.~Messaris, A.~S. Demirkol, R.~Tetzlaff, and L.~O. Chua.
\newblock The state change per cycle map: a novel system-theoretic analysis tool for periodically-driven reram cells.
\newblock {\em Frontiers in Electronic Materials}, 3, 2023.

\bibitem{7557034}
Alon Ascoli, Ronald Tetzlaff, and Leon~O. Chua.
\newblock The first ever real bistable memristors—part i: Theoretical insights on local fading memory.
\newblock {\em IEEE Transactions on Circuits and Systems II: Express Briefs}, 63(12):1091--1095, 2016.

\bibitem{Ascoli2016}
Alon Ascoli, Ronald Tetzlaff, and Leon~O. Chua.
\newblock {The first ever real bistable memristors - Part II: Design and analysis of a local fading memory system}.
\newblock {\em IEEE Transactions on Circuits and Systems II: Express Briefs}, 63(12), 2016.

\bibitem{Kumar2018}
Suhas Kumar and R.~Stanley Williams.
\newblock {Separation of current density and electric field domains caused by nonlinear electronic instabilities}.
\newblock {\em Nature Communications}, 2018.

\bibitem{Kumar2017b}
Suhas Kumar, John~Paul Strachan, and R.~Stanley Williams.
\newblock {Chaotic dynamics in nanoscale NbO2 Mott memristors for analogue computing}.
\newblock {\em Nature}, 2017.

\bibitem{Kumar2017a}
Suhas Kumar et~al.
\newblock {Physical origins of current and temperature controlled negative differential resistances in NbO2}.
\newblock {\em Nature Communications}, 2017.

\bibitem{Gibson2016}
Gary~A. Gibson et~al.
\newblock {An accurate locally active memristor model for S-type negative differential resistance in NbOx}.
\newblock {\em Applied Physics Letters}, 2016.

\bibitem{Demirkol2022}
Ahmet~Samil Demirkol, Ioannis Messaris, Alon Ascoli, and Ronald Tetzlaff.
\newblock {Pattern Formation in an M-CNN Structure Utilizing a Locally Active NbOx Memristor}.
\newblock In {\em Memristor Computing Systems}. 2022.

\bibitem{Nandi2019}
Sanjoy~Kumar Nandi, Shimul~Kanti Nath, Assaad~E. El-Helou, Shuai Li, Xinjun Liu, Peter~E. Raad, and Robert~G. Elliman.
\newblock {Current Localization and Redistribution as the Basis of Discontinuous Current Controlled Negative Differential Resistance in NbOx}.
\newblock {\em Advanced Functional Materials}, 29(50), 2019.

\bibitem{Li2019}
Shuai Li, Xinjun Liu, Sanjoy~Kumar Nandi, Shimul~Kanti Nath, and Robert~Glen Elliman.
\newblock {Origin of Current-Controlled Negative Differential Resistance Modes and the Emergence of Composite Characteristics with High Complexity}.
\newblock {\em Advanced Functional Materials}, 29(44), 2019.

\bibitem{Kumar2017}
Suhas Kumar, Noraica Davila, Ziwen Wang, Xiaopeng Huang, John~Paul Strachan, David Vine, A.~L. {David Kilcoyne}, Yoshio Nishi, and R.~{Stanley Williams}.
\newblock {Spatially uniform resistance switching of low current, high endurance titanium-niobium-oxide memristors}.
\newblock {\em Nanoscale}, 9(5), 2017.

\bibitem{9181036}
I.~Messaris, R.~Tetzlaff, A.~Ascoli, R.~S. Williams, S.~Kumar, and L.~Chua.
\newblock A simplified model for a nbo2 mott memristor physical realization.
\newblock In {\em 2020 IEEE International Symposium on Circuits and Systems (ISCAS)}, pages 1--5, 2020.

\bibitem{Chua2015}
Leon Chua.
\newblock {Everything you wish to know about memristors but are afraid to ask}.
\newblock {\em Radioengineering}, 2015.

\bibitem{6739164}
Deyan Lin, S.~Y.~Ron Hui, and Leon~O. Chua.
\newblock Gas discharge lamps are volatile memristors.
\newblock {\em IEEE Transactions on Circuits and Systems I: Regular Papers}, 61(7):2066--2073, 2014.

\bibitem{Adhikari2013}
Shyam~Prasad Adhikari, Maheshwar~Pd Sah, Hyongsuk Kim, and Leon~O. Chua.
\newblock {Three fingerprints of memristor}.
\newblock {\em IEEE Transactions on Circuits and Systems I: Regular Papers}, 2013.

\bibitem{Brown2022b}
Timothy~D Brown, Suhas Kumar, and R~Stanley Williams.
\newblock {Physics-based compact modeling of electro-thermal memristors: Negative differential resistance, local activity, and non-local dynamical bifurcations}.
\newblock {\em Applied Physics Reviews}, 9(1):11308, feb 2022.

\end{thebibliography}

\end{document}